\documentclass[11pt]{article}
\usepackage{epsfig} 
\newcommand{\sect}[1]{ \section{#1} \setcounter{equation}{0} }

\newcommand{\Dslash}{D \! \! \! \! /} 
 
\newcommand{\pslash}{p \! \! \! /}

\newcommand{\half}{\mbox{\small{$\frac{1}{2}$}}} 
 
\newcommand{\Nf}{N_{\!f}} 
\newcommand{\NA}{N_{\!A}} 
\newcommand{\NF}{N_{\!F}} 
\newcommand{\MSbar}{\overline{\mbox{MS}}} 

\begin{document}
\title{One loop gluon form factor and freezing of $\alpha_s$ in the 
Gribov-Zwanziger QCD Lagrangian}
\author{J.A. Gracey, \\ Theoretical Physics Division, \\ 
Department of Mathematical Sciences, \\ University of Liverpool, \\ P.O. Box 
147, \\ Liverpool, \\ L69 3BX, \\ United Kingdom.} 
\date{} 
\maketitle 
\vspace{5cm} 
\noindent 
{\bf Abstract.} We use the Gribov-Zwanziger Lagrangian in QCD to evaluate the 
one loop correction to the gluon propagator as a function of the Gribov volume 
and verify that the propagator vanishes in the infrared limit. Using the 
corresponding correction to the Faddeev-Popov ghost propagator we construct the 
renormalization group invariant coupling constant,
$\alpha^{\mbox{\footnotesize{eff}}}_S (p^2)$, from the gluon and ghost form 
factors and verify, using the Gribov mass gap condition, that it freezes 
out at zero momentum to a non-zero value. This is qualitatively consistent with
other approaches. We also show that there is an enhancement of the propagator
of one of the Zwanziger ghosts at two loops similar to that which occurs for 
the Faddeev-Popov ghosts. From the exact evaluation of the form factors we 
examine power corrections for the gluon propagator and the effective coupling. 
We find that both have the same qualitative behaviour in that the leading power
correction is $O(1/p^2)$ and not $O(1/(p^2)^2)$. 

\vspace{-19cm}
\hspace{13.5cm}
{\bf LTH-702} 

\newpage 

\sect{Introduction.}

In the late seventies, Gribov pointed out that fixing a gauge in a non-abelian
gauge theory was a non-trivial exercise in comparison with an abelian theory,
\cite{1}. In particular the gauge field was not fixed uniquely since for some
gauge fields one could construct at least one other copy which could be 
obtained from the first by a gauge transformation. To understand this problem 
and the relation to confinement, Gribov restricted the region of integration in
the path integral defining the non-abelian field theory to the region bounded 
by the {\em first} zero of the Faddeev-Popov operator, \cite{1}. The boundary 
of this region, known as the Gribov horizon, defines the Gribov volume and is 
parametrised by the quantity $\gamma$, of mass dimension one, known as the 
Gribov mass. Though inside the Gribov horizon one can still have gauge copies, 
\cite{2,3,4}. However, within this horizon is the fundamental modular region 
where each gauge field is uniquely defined, without copies, upon gauge fixing. 
Moreover, it had been suggested, \cite{5}, that any additional gauge copies 
inside the Gribov horizon do not affect the vacuum expectation values of any 
operator. Subsequent zeroes of the Faddeev-Popov operator beyond the first, 
define further regions of configuration space but it has been shown, \cite{1}, 
that they are equivalent to the first Gribov region which includes the origin. 
If one accepts that quantum chromodynamics (QCD), which is clearly a 
non-abelian gauge theory, has to be restricted to the volume defined by the 
Gribov horizon then Gribov argued that this had a profound effect on the 
infrared structure of the field theory, \cite{3,6,7,8,9,10,11}. For instance, 
in the Landau gauge Gribov showed that the ghost propagator diverged as 
$\frac{1}{(p^2)^2}$ as $p^2$~$\rightarrow$~$0$ rather than the usual 
$\frac{1}{p^2}$ behaviour which occurs in the ultraviolet (and perturbative) 
region, \cite{1}. This property, known as ghost enhancement, followed directly 
as a result of the Gribov mass satisfying a gap equation which was determined 
at one loop explicitly, \cite{1}. A comprehensive intoduction to the original 
Gribov article is provided in \cite{12}. 

In recent years, given the advance in non-perturbative techniques such as 
Dyson-Schwinger equations (DSE), the behaviour of the ghost propagator has been
studied in the Landau gauge in the infrared and is in agreement with Gribov's 
enhancement observation, \cite{13,14,15,16}. Though the actual power law 
exponent differs from that of Gribov, \cite{13,14,15,16}. Further, such studies
have opened up the possibility of examining other infrared features. For 
instance, it is believed the gluon propagator does not diverge at zero 
momentum, as its ultraviolet form would suggest, but either tends to zero or a 
finite value. Whilst Gribov demonstrated the gluon propagator vanished at the 
tree or classical level in the Landau gauge, \cite{1}, there appears to be 
no definitive agreement which of the latter scenarios persist in the quantum
theory. For instance, a recent lattice study argues that the data imply a
finite non-zero value for the gluon propagator at zero momentum, \cite{17}. 
Whilst stated in these general terms it is worth bearing in mind that there are
technical considerations such as gauge variance which may result in different 
pictures, when viewed in different calculational approaches and renormalization 
schemes. Though we note that throughout, our remarks will always be concerned
with the Landau gauge. Equally controversial is the behaviour of the strong 
coupling constant $\alpha_S (p^2)$. Phenomenologically it is believed that 
$\alpha_S (p^2)$ freezes out at zero momentum to a finite non-zero value,
\cite{18,19,20,21,22,23,24,25,26,27,28}. Though as we will illustrate later 
what its precise value is appears to depend on the method used to extract it 
from experimental data as well as how one chooses to define it. More 
theoretical predictions based on properties of the underlying quantum field 
theory itself such as analyticity arguments or DSE, rather than the fits to 
experimental data, favour a value higher than those from data, 
\cite{29,30,31,32,33}. Again these comments need to be tempered with the caveat
that theoretical predictions are made as a consequence of certain 
approximations and in renormalization schemes which are mass dependent, which 
imply a gauge {\em dependent} coupling constant. However, as the quantity 
measured in DSE and lattice studies is a renormalization group invariant in the
Landau gauge, which is measured at zero momentum, the hope is that such 
approximations do not significantly affect the estimates. Though some lattice 
and DSE studies do not see a finite freezing but instead observe an effective 
coupling which vanishes at zero momentum, \cite{34,35,36}. Whether this is due 
to finite volume effects as argued in \cite{33} by DSE studies, but 
subsequently rebutted in \cite{37} is still not clear. Moreover, it would seem 
that one of the issues in this technical debate centres on the validity and 
application of the underlying Slavnov-Taylor identities to the fully 
non-perturbative r\'{e}gime which governs the infrared behaviour. Though it is 
worth noting that an effective coupling can be defined from any of the vertices
of the QCD Lagrangian and the infrared behaviour of each does not have to be 
the same.  

In the context of the Gribov approach to the infrared structure the advance
made by Zwanziger, \cite{7,8,10,11}, in reformulating the Gribov path integral 
and its inherent non-locality deriving from the limitation to the Gribov region
is significant. In particular Zwanziger constructed a localized Lagrangian 
which involved extra ghost fields (which we will refer to as Zwanziger ghosts 
in contradistinction to the Faddeev-Popov ghosts), which when eliminated by 
their equations of motion reproduces the Gribov formulation of the problem. As 
of nearly equal significance is the fact that this Lagrangian (which we will 
refer to as the Gribov-Zwanziger Lagrangian) is renormalizable, 
\cite{10,38,39}, and explicitly contains the Gribov mass $\gamma$. This former 
property has been analyzed through the algebraic renormalization machinery and 
the Slavnov-Taylor identities have been established, \cite{38,39}. 
Interestingly despite the presence of the extra fields and a mass, no new 
independent renormalization constants are required in the Landau gauge to 
render the Gribov-Zwanziger Lagrangian finite. Moreover, as is evident in QCD 
when massive quarks are present the $\MSbar$ renormalization of the coupling 
constant, gluon and Faddeev-Popov ghost fields remain unaffected by a non-zero
$\gamma$, \cite{38,39}. Crucially the general formalism implies that the 
Gribov-Zwanziger Lagrangian can be used for performing loop calculations. 
Indeed Zwanziger, \cite{8}, reproduced the one loop gap equation of \cite{1} by
evaluating the vacuum expectation value of the Gribov mass operator which is 
equivalent to the implementation of the Gribov horizon condition. More recently
this has been examined at two loops in \cite{40} where the two loop correction 
to Gribov's gap equation has been determined in the Landau gauge with massless 
quarks in the $\MSbar$ scheme. As a check that the gap equation is consistent, 
the two loop correction to the Faddeev-Popov ghost propagator was computed and 
shown that the ghost propagator enhancement is preserved as a 
$\frac{1}{(p^2)^2}$ behaviour precisely as a result of the two loop gap 
equation, \cite{40}. 

Thus armed with a renormalizable localized Lagrangian, which at two loops 
appears to reproduce the observed behaviour noted by Gribov, we can attack
the problem of the behaviour of other quantities in the infrared which we have
discussed above. This is the aim of this article where we will compute the one
loop correction to the gluon, Gribov-Zwanziger and Faddeev-Popov ghost
propagators exactly. As will be evident this is not the straightforward 
exercise it would normally be in the absence of the Gribov-Zwanziger ghosts or
the Gribov mass. However, we will be able to study the $p^2$~$\rightarrow$~$0$ 
behaviour of the gluon propagators and see if it vanishes at the one loop level
or not. Equally worth studying is the effective coupling constant freezing at 
zero momentum. Since lattice and DSE studies examine an effective coupling 
constant which is renormalization group invariant but which depends only on the
gluon and Faddeev-Popov ghost form factors, we will be able to construct this 
quantity and then take the $p^2$~$\rightarrow$~$0$ limit. We believe this will 
be constructive for reasons beyond the qualitative calculation we will perform. 
Specifically, previous studies have primarily been numerical and deduced either 
from experiment or by solving QCD with lattice or DSE techniques. The actual 
mechanism and which of the quantum fields truly drive the freezing have not 
really been studied. Having a Lagrangian where the calculations can be 
constructed explicitly will, we believe, be important in furthering that 
understanding. Given that we evaluate the propagators as {\em exact} functions 
of $\gamma$ and the momentum, we can also study the corrections to the 
effective coupling in the alternative limit of $\gamma^2$~$\rightarrow$~$0$. 
The motivation for this is to examine whether one can not only produce power 
corrections but also to see whether such corrections are $O(1/p^2)$ as 
suggested by the lattice computation of \cite{43,44} or $O(1/(p^2)^2)$ as would
be expected on the grounds of a condensation of the gauge invariant operator 
$G^a_{\mu\nu} G^{a\,\mu\nu}$. Finally, in referring to the work we present as 
qualitative, it is worth briefly clarifying at the outset that what we mean by 
this is that performing {\em one} loop calculations, where behaviour in the 
infrared appears to be consistent with non-perturbative highly intensive DSE 
and lattice studies, our analysis should only be regarded as comparative. 

The paper is organised as follows. In section $2$ we recall the main structure
of the Gribov-Zwanziger Lagrangian and discuss the propagators and gap 
equation. The formal background to the computation of the one loop corrections
to the $2$-point functions of the Gribov-Zwanziger Lagrangian is introduced
in section $3$ prior to giving the details of the explicit computation in 
section $4$. Section $5$ is devoted to the problem of the freezing of the 
effective coupling constant in the $p^2$~$\rightarrow$~$0$ limit whilst we 
discuss the structure of the propagators and effective coupling constant in the
other limit $\gamma^2$~$\rightarrow$~$0$ in section $6$. Finally, our 
conclusions are given in section $7$ and an appendix collects the full 
expressions for the exact $2$-point functions. 

\sect{Gribov-Zwanziger Lagrangian.}

We begin by defining the localized Lagrangian derived by Zwanziger, \cite{8},
which replaces the non-local formulation of the gauge fixing condition of 
Gribov, \cite{1}, concentrating on the Landau gauge. In addition to the usual
Faddeev-Popov ghosts, $c^a$ and $\bar{c}^a$, of the canonical gauge fixing
construction, the Gribov-Zwanziger Lagrangian involves additional Zwanziger
ghosts. These fields, denoted by $\{\phi^{ab}_\mu,\bar{\phi}^{ab}_\mu\}$ and
$\{\omega^{ab}_\mu,\bar{\omega}^{ab}_\mu\}$, are spin-$1$ and carry two colour
indices. The latter set are anticommuting Grassmann fields whilst the former
are commuting. Thus the full Gribov-Zwanziger Lagrangian is, \cite{8}, 
\begin{eqnarray}
L^{\mbox{\footnotesize{GZ}}} &=& L^{\mbox{\footnotesize{QCD}}} ~+~ 
\bar{\phi}^{ab \, \mu} \partial^\nu \left( D_\nu \phi_\mu \right)^{ab} ~-~ 
\bar{\omega}^{ab \, \mu} \partial^\nu \left( D_\nu \omega_\mu \right)^{ab} 
\nonumber \\  
&& -~ g f^{abc} \partial^\nu \bar{\omega}^{ae}_\mu \left( D_\nu c \right)^b
\phi^{ec \, \mu} ~-~ \frac{\gamma^2}{\sqrt{2}} \left( f^{abc} A^{a \, \mu} 
\phi^{bc}_\mu ~+~ f^{abc} A^{a \, \mu} \bar{\phi}^{bc}_\mu \right) ~+~ 
\frac{d \NA \gamma^4}{2g^2} 
\label{laggz}
\end{eqnarray} 
with 
\begin{equation} 
L^{\mbox{\footnotesize{QCD}}} ~=~ \frac{1}{4} G_{\mu\nu}^a 
G^{a \, \mu\nu} ~-~ \frac{1}{2\alpha} (\partial^\mu A^a_\mu)^2 ~-~ 
\bar{c}^a \partial^\mu D_\mu c^a ~+~ i \bar{\psi}^{iI} \Dslash \psi^{iI} 
\end{equation} 
where $A^a_\mu$ is the gluon, $\psi^{iI}$ is the (massless) quark and the 
indices lie in the ranges $1$~$\leq$~$a$~$\leq$~$\NA$, 
$1$~$\leq$~$i$~$\leq$~$\NF$ and $1$~$\leq$~$I$~$\leq$~$\Nf$ where $\NF$ and
$\NA$ are the respective dimensions of the fundamental and adjoint 
representations and $\Nf$ is the number of quark flavours. The covariant
derivatives are given by  
\begin{eqnarray} 
D_\mu c^a &=& \partial_\mu c^a ~-~ g f^{abc} A^b_\mu c^c ~~,~~ 
D_\mu \psi^{iI} ~=~ \partial_\mu \psi^{iI} ~+~ i g T^a A^a_\mu \psi^{iI} 
\nonumber \\
\left( D_\mu \phi_\nu \right)^{ab} &=& \partial_\mu \phi^{ab}_\nu ~-~
g f^{acd} A^c_\mu \phi^{db}_\nu ~. 
\end{eqnarray} 
For completeness, we have included the covariant gauge fixing parameter 
$\alpha$ which is in principle required in the derivation of the gluon 
propagators. The Gribov mass, $\gamma$, is present in the $2$-point term mixing
between the gluon and the Zwanziger ghosts. In various articles by other 
authors the factor of $\frac{1}{\sqrt{2}}$ is absent. However, as we will 
observe later this choice is essential to retaining a gluon propagator which is
the same form as that originally derived by Gribov, \cite{1}. Though, we note 
now that changing the numerical coefficient of $\gamma$ does not alter the 
overall physics but merely rescales the value of the gluon mass as a function 
of $\gamma$. The presence of the mixed $2$-point term is problematic from the 
point of view of developing Feynman rules since one would have a mixed gluon
Zwanziger ghost propagator. One way to circumvent this would be to redefine 
the $A^a_\mu$ and $\phi^{ab}_\mu$ fields in such a way that the $2$-point 
term is absent. It transpires that this is cumbersome to implement and in 
earlier work, \cite{40}, an approach has been found which allows one to
systematically evaluate Feynman diagrams with a mixed propagator and to derive
the two loop correction to the Gribov mass gap. 

To derive the Feynman rules from (\ref{laggz}) for the vertices of the 
interaction Lagrangian is straightforward. However, the main difficulty is to
determine the propagator for the gluon in the presence of the mixing term.
Although this was described in \cite{40} it is instructive to develop the
derivation since it serves as the basis for deducing the one loop correction
to the gluon form factor. First, we note that in general in deriving the
propagators from the quadratic part of a Lagrangian the fields are first
transformed to momentum space. Therefore, concentrating only on the gluon and
Zwanziger ghost part of (\ref{laggz}) we have 
\begin{eqnarray}
L^{\mbox{\footnotesize{GZ}}}_{\mbox{\footnotesize{quad}}} &=& \frac{1}{2} 
A^a_\mu(-p) \left[ p^2 \eta^{\mu\nu} - \left( 1 - \frac{1}{\alpha} \right) 
p^\mu p^\nu \right] A^a_\nu(p) ~-~ \frac{\gamma^2}{\sqrt{2}} f^{abc} 
A^a_\mu(-p) \phi^{bc \, \mu} (p) \nonumber \\
&& +~ \frac{\gamma^2}{\sqrt{2}} f^{abc} \bar{\phi}^{ab\,\mu}(-p) 
A^c_\mu(p) ~-~ p^2 \bar{\phi}^{ab\,\mu}(-p) \phi^{ab}_\mu(p) ~. 
\end{eqnarray} 
If $\gamma$ was zero then the inversion of the momentum space operator 
associated with the terms quadratic in the fields proceeds in a textbook
manner. However, for $\gamma$ non-zero we must first write 
$L^{\mbox{\footnotesize{GZ}}}_{\mbox{\footnotesize{quad}}}$ in matrix form with
respect to the basis $\left\{ \frac{1}{\sqrt{2}} A^a_\mu, \phi^{ab}_\mu 
\right\}$. The inclusion of the factor $\frac{1}{\sqrt{2}}$ into the basis is 
crucial since $L^{\mbox{\footnotesize{GZ}}}_{\mbox{\footnotesize{quad}}}$ 
clearly now becomes  
\begin{equation}
\left( \frac{A^a_\mu(-p)}{\sqrt{2}}, \bar{\phi}^{ab}_\mu(-p) \right)
\left(
\begin{array}{cc}
\left[ p^2 \eta^{\mu\nu} - \left( 1 - \frac{1}{\alpha} \right) p^\mu p^\nu
\right] \delta^{ac} &
- \gamma^2 f^{acd} \eta^{\mu\nu} \\
- \gamma^2 f^{cab} \eta^{\mu\nu} &  
- p^2 \eta^{\mu\nu} \delta^{ac} \delta^{bd} \\
\end{array}
\right)
\left(
\begin{array}{c}
\frac{A^c_\nu(p)}{\sqrt{2}} \\ \phi^{cd}_\nu(p) \\
\end{array}
\right) ~. 
\label{qdmat} 
\end{equation} 
Inverting the matrix using the unit matrix 
\begin{equation} 
\left(
\begin{array}{cc}
\eta_{\mu\nu} \delta^{ab} & 0 \\ 0 & \eta_{\mu\nu}
\delta^{ac} \delta^{bd} \\ 
\end{array} \right) 
\end{equation}
then the $\alpha$ dependent propagators are 
\begin{eqnarray}
\langle A^a_\mu(p) A^b_\nu(-p) \rangle &=& -~ 
\delta^{ab} p^2 \left[ \frac{P_{\mu\nu}(p)}{[(p^2)^2+C_A\gamma^4]} ~+~ 
\frac{\alpha L_{\mu\nu}(p)}{[(p^2)^2+\alpha C_A\gamma^4]} \right] \nonumber \\  
\langle A^a_\mu(p) \bar{\phi}^{bc}_\nu(-p) \rangle &=& -~ 
\frac{f^{abc}\gamma^2}{\sqrt{2}} \left[ 
\frac{P_{\mu\nu}(p)}{[(p^2)^2+C_A\gamma^4]} ~+~ 
\frac{\alpha L_{\mu\nu}(p)}{[(p^2)^2+\alpha C_A\gamma^4]} \right] \nonumber \\  
\langle \phi^{ab}_\mu(p) \bar{\phi}^{cd}_\nu(-p) \rangle &=& -~ 
\frac{\delta^{ac}\delta^{bd}}{p^2}\eta_{\mu\nu} \nonumber \\
&& +~ \frac{f^{abe}f^{cde}\gamma^4}{p^2} \left[ 
\frac{P_{\mu\nu}(p)}{[(p^2)^2+C_A\gamma^4]} ~+~ 
\frac{\alpha L_{\mu\nu}(p)}{[(p^2)^2+\alpha C_A\gamma^4]} \right]  
\label{gzpropal} 
\end{eqnarray} 
where 
\begin{equation}
P_{\mu\nu}(p) ~=~ \eta_{\mu\nu} ~-~ \frac{p_\mu p_\nu}{p^2} ~~~,~~~  
L_{\mu\nu}(p) ~=~ \frac{p_\mu p_\nu}{p^2} 
\label{projs} 
\end{equation}
are the respective transverse and longitudinal projectors and we have been 
careful to restore the basis vector $\left\{ \frac{1}{\sqrt{2}} A^a_\mu, 
\phi^{ab}_\mu \right\}$ and its transpose prior to deducing (\ref{gzpropal}) 
which is the origin of the location of the factor of $\frac{1}{\sqrt{2}}$ in 
the mixed propagator. Further, without the factor of $\frac{1}{\sqrt{2}}$ in 
the mixed term of (\ref{laggz}) one would not only have $\sqrt{2}\gamma^2$ in 
the off-diagonal terms of the matrix of (\ref{qdmat}) but also one would have a
common factors of $[(p^2)^2 + 2 C_A \gamma^4]$ and $[(p^2)^2 + 2 \alpha C_A 
\gamma^4]$ in the propagators. In the Landau gauge calculation of \cite{1} this
additional numerical factor of $2$ was absent and we choose to be consistent 
with \cite{1} as our convention. Moreover, as we will later have to write 
$[(p^2)^2 + C_A \gamma^4]$ as the product of the more conventional factors 
$[p^2 + i \sqrt{C_A} \gamma^2]$ and $[p^2 - i \sqrt{C_A} \gamma^2]$ the choice 
in (\ref{laggz}) therefore also turns out to be the most convenient. Next we 
note that in the Landau gauge the final propagators for the gluon and Zwanziger
ghosts are, \cite{8},  
\begin{eqnarray}
\langle A^a_\mu(p) A^b_\nu(-p) \rangle &=& -~ 
\frac{\delta^{ab}p^2}{[(p^2)^2+C_A\gamma^4]} P_{\mu\nu}(p) \nonumber \\  
\langle A^a_\mu(p) \bar{\phi}^{bc}_\nu(-p) \rangle &=& -~ 
\frac{f^{abc}\gamma^2}{\sqrt{2}[(p^2)^2+C_A\gamma^4]} P_{\mu\nu}(p) 
\nonumber \\  
\langle \phi^{ab}_\mu(p) \bar{\phi}^{cd}_\nu(-p) \rangle &=& -~ 
\frac{\delta^{ac}\delta^{bd}}{p^2}\eta_{\mu\nu} ~+~  
\frac{f^{abe}f^{cde}\gamma^4}{p^2[(p^2)^2+C_A\gamma^4]} P_{\mu\nu}(p) ~. 
\label{gzprop}
\end{eqnarray} 
Whilst this may be an unusual way of considering the quadratic terms we note
that it is primarily forced, aside from the original mixing, by the fact that 
the gluon is a real field whereas the Zwanziger ghosts are complex. Finally, 
for completeness we note that the propagators of the remaining fields are
\begin{eqnarray}  
\langle c^a(p) \bar{c}^b(-p) \rangle &=& -~ \frac{\delta^{ab}}{p^2} ~~~,~~~ 
\langle \psi^{iI}(p) \bar{\psi}^{jJ}(-p) \rangle ~=~ 
i\delta^{ij}\delta^{IJ} \frac{\pslash}{p^2} \nonumber \\ 
\langle \omega^{ab}_\mu(p) \bar{\omega}^{cd}_\nu(-p) \rangle &=& -~ 
\frac{\delta^{ac}\delta^{bd}\eta_{\mu\nu}}{p^2} ~.  
\end{eqnarray}  

Having summarized the derivation of the mixed propagators we note that the
$\phi^{ab}_\mu$ fields are the ones which implement the Gribov horizon
condition, \cite{8}. To recap from (\ref{laggz}) the equation of motion of
$\bar{\phi}^{ab}_\mu$ gives  
\begin{equation}
\phi^{ab}_\mu ~=~ \frac{\gamma^2}{\sqrt{2}} f^{abc} 
\frac{1}{\partial^\nu D_\nu} A^c_\mu
\end{equation}
whence one has 
\begin{equation}
f^{abc} \langle A^{a \, \mu}(x) \phi^{bc}_\mu(x) \rangle ~=~ 
\frac{d \NA \gamma^2}{\sqrt{2}g^2} 
\label{gapdef}
\end{equation} 
where $\gamma$ is defined by the horizon condition, \cite{8}, 
\begin{equation}
\left\langle A^a_\mu \, \frac{1}{\partial^\nu D_\nu} A^{a \, \mu} 
\right\rangle ~=~ \frac{d \NA}{C_A g^2} ~.  
\end{equation} 
In noting this then it is elementary to reproduce Gribov's original gap 
equation by integrating the mixed propagator over the loop momentum $p$ of
(\ref{gzprop}) to give the $\MSbar$ result, \cite{1,8},  
\begin{equation} 
1 ~=~ C_A \left[ \frac{5}{8} ~-~ \frac{3}{8} \ln \left( 
\frac{C_A\gamma^4}{\mu^4} \right) \right] a ~+~ O(a^2) 
\label{gap1} 
\end{equation} 
where $a$ $=$ $g^2/(16\pi^2)$ and $\mu$ is the renormalization scale introduced
to ensure that the coupling constant remains dimensionless in $d$ dimensions
when dimensional regularization is used. We have also absorbed the usual factor
of $4\pi e^{-\bar{\gamma}}$ into $\mu^2$ where $\bar{\gamma}$ is the 
Euler-Mascheroni constant. Although the gap equation is divergent the 
infinities are absorbed in the renormalization constants defined by  
\begin{eqnarray} 
A^{a \, \mu}_{\mbox{\footnotesize{o}}} &=& \sqrt{Z_A} \, A^{a \, \mu} ~~,~~ 
c^a_{\mbox{\footnotesize{o}}} ~=~ \sqrt{Z_c} \, c^a ~~,~~ 
\phi^{ab}_{\mu \, \mbox{\footnotesize{o}}} ~=~ \sqrt{Z_\phi} \, 
\phi^{ab}_\mu ~~,~~ 
\omega^{ab}_{\mu \, \mbox{\footnotesize{o}}} ~=~ \sqrt{Z_\omega} \, 
\omega^{ab}_\mu \nonumber \\ 
\psi_{\mbox{\footnotesize{o}}} &=& \sqrt{Z_\psi} \psi ~~,~~ 
g_{\mbox{\footnotesize{o}}} ~=~ Z_g \, g ~~,~~ 
\gamma_{\mbox{\footnotesize{o}}} ~=~ Z_\gamma \, \gamma 
\label{rencon} 
\end{eqnarray} 
where ${}_{\mbox{\footnotesize{o}}}$ denotes a bare quantity and $g$ and
$\gamma$ are the running coupling constant and running Gribov mass
respectively. We note that it was shown in \cite{38,39} that in the Landau 
gauge
\begin{equation}
Z_c ~=~ Z_\phi ~=~ Z_\omega ~=~ \frac{1}{Z_g \sqrt{Z_A}}
\end{equation} 
and, in our conventions,  
\begin{equation} 
Z_\gamma ~=~ \left( Z_A Z_c \right)^{-1/4} ~. 
\end{equation}
Explicitly at three loops one has in $\MSbar$ 
\begin{eqnarray} 
\gamma_\gamma(a) &=& \left[ 16 T_F \Nf - 35 C_A \right] \frac{a}{48} ~+~ 
\left[ 280 C_A T_F \Nf - 449 C_A^2 + 192 C_F T_F \Nf \right]
\frac{a^2}{192} \nonumber \\  
&& +~ \left[ ( 486 \zeta(3) - 75607 ) C_A^3 + ( 89008 - 15552 \zeta(3) )
C_A^2 T_F \Nf \right. \nonumber \\
&& \left. ~~~~+~ ( 20736 \zeta(3) + 19920 ) C_A C_F T_F \Nf - 12352 C_A T_F^2
\Nf^2 \right. \nonumber \\
&& \left. ~~~~-~ 3456 C_F^2 T_F \Nf - 8448 C_F T_F^2 \Nf^2 \right] 
\frac{a^3}{6912} ~+~ O(a^4) ~. 
\end{eqnarray} 
Though the four loop $\MSbar$ expression for the colour group $SU(N)$ is 
available in \cite{45}, from knowledge of the Landau gauge gluon and 
Faddeev-Popov ghost anomalous dimensions. 

In \cite{1} the gap equation was used to show that in the infrared the ghost
propagator is enhanced and behaves as $\frac{1}{(p^2)^2}$ as 
$p^2$~$\rightarrow$~$0$. This was obtained by computing the single one loop
correction to the ghost $2$-point function and noting that in the $p^2$
expansion of the diagram the leading $p^2$ correction was precisely the 
$O(a)$ part of (\ref{gap1}) for all $\mu$. More recently (\ref{laggz}) has
been used to compute the two loop correction to the horizon condition,
(\ref{gapdef}), and the ghost $2$-point function. In \cite{40} the $\MSbar$
correction for massless quarks, is 
\begin{eqnarray} 
1 &=& C_A \left[ \frac{5}{8} - \frac{3}{8} \ln \left( 
\frac{C_A\gamma^4}{\mu^4} \right) \right] a \nonumber \\ 
&& +~ \left[ C_A^2 \left( \frac{2017}{768} - \frac{11097}{2048} s_2
+ \frac{95}{256} \zeta(2)
- \frac{65}{48} \ln \left( \frac{C_A\gamma^4}{\mu^4} \right)
+ \frac{35}{128} \left( \ln \left( \frac{C_A\gamma^4}{\mu^4} \right)
\right)^2 \right. \right. \nonumber \\
&& \left. \left. ~~~~~~~~~~~~+~ \frac{1137}{2560} \sqrt{5} \zeta(2) 
- \frac{205\pi^2}{512} \right) \right. \nonumber \\
&& \left. ~~~~~+~ C_A T_F \Nf \left( -~ \frac{25}{24} - \zeta(2)
+ \frac{7}{12} \ln \left( \frac{C_A\gamma^4}{\mu^4} \right)
- \frac{1}{8} \left( \ln \left( \frac{C_A\gamma^4}{\mu^4} \right) \right)^2 
+ \frac{\pi^2}{8} \right) \right] a^2 \nonumber \\
&& +~ O(a^3)  
\label{gap2} 
\end{eqnarray} 
where $s_2$ $=$ $(2\sqrt{3}/9) \mbox{Cl}_2(2\pi/3)$, $\mbox{Cl}_2(x)$ is the
Clausen function and $\zeta(n)$ is the Riemann zeta function. Interestingly
the $O(p^2)$ part of the $O(a^2)$ correction to the ghost $2$-point function  
matched precisely with the $O(a^2)$ part of (\ref{gap2}) for all $\mu$ so that
in (\ref{laggz}) the ghost propagator is still enhanced in the infrared limit.
This is one of the conditions of the Kugo-Ojima confinement mechanism,
\cite{46}, which ensures that the Gribov-Zwanziger Lagrangian is an important 
tool for exploring the infrared structure of QCD from an analytic point of 
view.

In the context of this Faddeev-Popov ghost enhancement at two loops in
(\ref{laggz}) we have also studied the structure of the $2$-point function of 
the anticommuting Zwanziger ghost field $\omega^{ab}_\mu$ in the same infrared 
limit. Although this is a spin-$1$ field it has some similarity with the 
corresponding Faddeev-Popov ghost field $c^a$ since they both have the same 
wave function renormalization constant. It transpires that in evaluating the
$1$ one loop and $31$ two loop Feynman diagrams contributing to the 
$\omega^{ab}_\mu$ $2$-point function the same phenomenon occurs as the 
Faddeev-Popov ghost. The two loop gap equation, (\ref{gap2}), emerges in the 
same way in the $p^2$ expansion of the $2$-point function so that the 
$\omega^{ab}_\mu$ ghost is also enhanced in the infrared. In other words as 
$p^2$~$\rightarrow$~$0$ the propagator behaves as $\frac{1}{(p^2)^2}$ rather 
than having the $\frac{1}{p^2}$ behaviour of the initial propagator. It is 
worth remarking that as far as we are aware the discussion of \cite{46} centred
on a theory which involved only Faddeev-Popov ghosts and not the extension with
the extra Zwanziger ghosts. Whilst in principle one would have to revisit that 
analysis to establish that the Kugo-Ojima mechanism still holds for 
(\ref{laggz}), the fact that the $\omega^{ab}_\mu$ propagator has precisely the
same structure of the Faddeev-Popov ghost would seem to imply that the 
Kugo-Ojima confinement criterion is still valid for (\ref{laggz}). Indeed it is
worth observing that in the specific calculation, from the form of the 
$\omega^{ab}_\mu$ propagator and its interaction with the gluon, the Lorentz 
structure of the $\omega^{ab}_\mu$ $2$-point function actually factors off and 
effectively leaves the Faddeev-Popov ghost $2$-point function.  

\sect{Formal derivation of propagators.}

We now turn to the computation of the one loop corrections to the gluon
propagator. However, given the nature of the problem we will also deduce 
information on the mixed propagator and $\phi^{ab}_\mu$ propagator 
simultaneously. First, we recall the normal procedure to determine loop 
corrections to a field where there is no $2$-point mixing. Essentially one 
computes the corrections to the $2$-point function and then inverts this 
expression truncating at the order in the coupling constant one is interested 
in. For (\ref{laggz}) the procedure is the same. However, one is dealing with a
$2$~$\times$~$2$ matrix and has to first determine the corrections to each 
$2$-point function which comprise the elements of the matrix. Once this has 
been achieved then the matrix can be inverted and all the propagators deduced 
from the appropriate elements. Several comments are worth making at this point.
If one considers the location of the gluon propagator in the inverted matrix of
$2$-point functions its structure is derived from two pieces. One is the 
determinant of the matrix which by definition involves all the elements of the 
original matrix and which will contribute to the denominator of the gluon 
propagator. The other piece will be located in the numerator of the final 
propagator correction and will also involve pieces from the Zwanziger ghost 
$\phi^{ab}_\mu$ $2$-point function. In other words the fields implementing the 
horizon condition will form a central, and as will be seen later, a crucial 
role in the behaviour of the gluon propagator and effective coupling constant 
in the infrared. It is worth recalling that these properties only result as a 
consequence of a non-zero $\gamma$ and hence the presence of a mixing term. 

We have described our strategy in a general formal way at the outset since the
practicalities of the inversion is technical. First, one is dealing with 
spin-$1$ fields which require a gauge fixing term to have a non-singular
inversion in momentum space. As this is standard we note that one can either
introduce a gauge fixing term as we did in the previous section before setting
$\alpha$~$=$~$0$ to specify the Landau gauge we are interested in, or one can
work in the Landau gauge directly and factor off the common transverse 
projector $P_{\mu\nu}(p)$ from the $2$-point matrix. We choose to do the
latter. However, another complication of the inversion lies in correctly taking
account of the group theory structure of the $\phi^{ab}_\mu$ $2$-point function
corrections. For the gluon itself and the mixed $2$-point functions the one
loop corrections have the same group structure as the Lagrangian term. However,
as the $\phi^{ab}_\mu$-field carries two group indices then its $2$-point
function will involve four colour index objects. If we denote these indices
by the set $\{a,b,c,d\}$ then we take the basis of independent rank four
objects which can arise from the one loop diagrams as
\begin{equation}
\left\{ \delta^{ac} \delta^{bd}, \delta^{ad} \delta^{bc}, 
\delta^{ab} \delta^{cd}, f^{ace} f^{bde}, f^{abe} f^{cde}, d_A^{abcd} 
\right\}
\end{equation}
where the Jacobi identity excludes $f^{ade} f^{bce}$ from being independent.
Though we could equally have chosen either of the other two combinations. The 
object $d_A^{abcd}$ is defined as  
\begin{equation}
d_A^{abcd} ~=~ \frac{1}{6} \mbox{Tr} \left( T_A^a T_A^{(b} T_A^c T_A^{d)} 
\right)
\end{equation} 
where $\left( T^a \right)_{bc}$ $=$ $-$ $i f^{abc}$ is the adjoint 
representation of the group generators and $d_R^{abcd}$ is the totally 
symmetric trace of four group generators in the $R$ representation, \cite{47}. 
To proceed with the inversion of the mixed $2$-point function this colour 
structure has to be taken into account in the $\phi^{ab}_\mu$ $2$-point 
function and its corresponding inverse element. As this is a formal discussion 
we will not cloud the inversion by including the explicit results of the loop 
integrals at this stage but take the general structure of the matrices we are 
interested in as follows. The matrix of $2$-point functions at one loop is  
\begin{equation}
\left(
\begin{array}{cc}
p^2 \delta^{ac} & - \gamma^2 f^{acd} \\
- \gamma^2 f^{cab} & - p^2 \delta^{ac} \delta^{bd} \\
\end{array}
\right) \, + \, 
\left(
\begin{array}{cc}
X \delta^{ac} & U f^{acd} \\
M f^{cab} & Q \delta^{ac} \delta^{bd} + W f^{ace} f^{bde} + R f^{abe} f^{cde}
+ S d_A^{abcd} \\
\end{array}
\right) a ~+~ O(a^2) 
\label{twoptdef} 
\end{equation} 
with respect to the basis $\left\{ \frac{1}{\sqrt{2}} A^a_\mu, \phi^{ab}_\mu 
\right\}$ where we have factored off the Lorentz structure. The quantities
$X$, $U$, $M$, $Q$, $W$, $R$ and $S$ represent the one loop corrections. 
Similarly the Landau gauge inverse will be of the form 
\begin{eqnarray}
&& \left(
\begin{array}{cc}
\frac{p^2}{[(p^2)^2+C_A\gamma^4]} \delta^{cp} & 
- \frac{\gamma^2}{[(p^2)^2+C_A\gamma^4]} f^{cpq} \\
- \frac{\gamma^2}{[(p^2)^2+C_A\gamma^4]} f^{pcd} & 
- \frac{1}{p^2} \delta^{cp} \delta^{dq} 
+ \frac{\gamma^4}{p^2[(p^2)^2+C_A\gamma^4]} f^{cdr} f^{pqr} \\
\end{array}
\right) \nonumber \\
&+& \left(
\begin{array}{cc}
A \delta^{cp} & C f^{cpq} \\
E f^{pcd} & G \delta^{cp} \delta^{dq} + J f^{cpe} f^{dqe} + K f^{cde} f^{pqe}
+ L d_A^{cdpq} \\
\end{array}
\right) a ~+~ O(a^2) 
\label{ffdef} 
\end{eqnarray} 
where we have included the propagators from the previous section and the 
quantities $A$, $C$, $E$, $G$, $J$, $K$ and $L$ will depend on the one loop
corrections defined in (\ref{twoptdef}). Given these forms it is 
straightforward to check that 
\begin{eqnarray} 
A &=& -~ \frac{1}{[(p^2)^2+C_A\gamma^4]^2} \left[ (p^2)^2 X - C_A \gamma^2 p^2
U - C_A \gamma^2 p^2 M + C_A \gamma^4 \left( Q + C_A R + \half C_A W \right)
\right] \nonumber \\  
C &=& \frac{1}{[(p^2)^2+C_A\gamma^4]^2} \left[ \gamma^2 p^2 X - C_A \gamma^4 M 
+ (p^2)^2 U - \gamma^2 p^2 \left( Q + C_A R + \half C_A W \right) \right] 
\nonumber \\  
E &=& \frac{1}{[(p^2)^2+C_A\gamma^4]^2} \left[ \gamma^2 p^2 X - C_A \gamma^4 U 
+ (p^2)^2 M - \gamma^2 p^2 \left( Q + C_A R + \half C_A W \right) \right] 
\nonumber \\  
G &=& -~ \frac{Q}{(p^2)^2} ~~~,~~~ J ~=~ -~ \frac{W}{(p^2)^2} ~~~,~~~
L ~=~ -~ \frac{S}{(p^2)^2} \nonumber \\ 
K &=& -~ \frac{1}{[(p^2)^2+C_A\gamma^4]^2} \left[ \gamma^4 X + \gamma^2 p^2 U 
+ \gamma^2 p^2 M + (p^2)^2 R - \gamma^4 \left( Q + \half C_A W \right) \right] 
\nonumber \\ 
&& +~ \ 
\frac{\gamma^4}{(p^2)^2[(p^2)^2+C_A\gamma^4]} \left[ Q + \half C_A W  \right] 
\label{propform}
\end{eqnarray} 
by ensuring that the $O(a)$ term of the product of (\ref{twoptdef}) and
(\ref{ffdef}) vanishes. As a check on this inversion we note that $C$~$=$~$E$
if $M$~$=$~$U$. Also in determining the $O(a)$ correction to the mixed 
propagator of (\ref{gzpropal}) $C$ and $E$ have to be divided by $\sqrt{2}$ as
was noted after (\ref{projs}). We have derived these corrections at the formal 
level to indicate how involved the one loop corrections to the propagators are.
However, given the intricate relation with the $2$-point function corrections 
it will transpire that they will only be of use at this formal level. 

\sect{One loop corrections.}

We now detail the computation of the one loop integrals. First, the required
Feynman diagrams are generated with the use of the {\sc Qgraf} package,
\cite{48}. For the gluon and $\phi^{ab}_\mu$ $2$-point functions there are 
eight and two diagrams respectively whilst there are two diagrams contributing 
to either of the mixed $2$-point functions giving a total of fourteen. Although
we are primarily interested in examining the $p^2$~$\rightarrow$~$0$ limit of 
the propagators to ascertain whether the one loop gluon propagator still 
vanishes as it does at one loop, we have decided to evaluate the one loop 
correction {\em exactly} as a function of $p^2$. This requires the explicit 
evaluation of the one type of master integral which is of the form
\begin{equation}
I_1(p,m_1^2,m_2^2;\alpha,\beta) ~=~ \int_k \frac{1}{[k^2+m_1^2]^\alpha  
[(k-p)^2+m_2^2]^\beta}
\label{intdef} 
\end{equation}  
where at one loop $\alpha$ and $\beta$ are integers and the mass arguments take
any combination of values in the set $\{0, i \sqrt{C_A} \gamma^2, - i
\sqrt{C_A} \gamma^2 \}$. This form, (\ref{intdef}), derives from rewriting the
propagators with the Gribov structure $p^2/[(p^2)+C_A\gamma^4]$ using partial
fractions such as
\begin{equation}
\frac{p^2}{[(p^2)^2 + C_A\gamma^4]} ~=~ \frac{1}{2} \left( 
\frac{1}{[p^2+i\sqrt{C_A}\gamma^2]} ~+~ \frac{1}{[p^2-i\sqrt{C_A}\gamma^2]} 
\right)
\label{parfra} 
\end{equation} 
producing standard propagators but with a width. However, it is worth noting 
that in any final expression we derive, the answer must be real given that the 
initial integrals are real which provides an internal check on the computation.
On a practical note we record that the decomposition (\ref{parfra}) is readily 
implemented in a computer algebra programme written in the symbolic 
manipulation language {\sc Form}, \cite{49}. Indeed the complete calculation we
describe here has been performed automatically with {\sc Form}.

Whilst the type of master integral, $I_1(p,m_1^2,m_2^2;1,1)$ has been studied
and exploited many times we note that the key difference here is the presence
of the complex mass. However, in the exact evaluation of our integrals we note
that we use the formal results for $I_1(p,m_1^2,m_2^2;1,1)$ with real $m_i^2$
before analytically continuing to the values $\pm i \sqrt{C_A} \gamma^2$ we are
interested in when the masses are non-zero. For completeness we note that the 
results for the two central integrals we use are, expanded to the finite parts, 
\begin{eqnarray}
I_1(p,i\sqrt{C_A}\gamma^2,i\sqrt{C_A}\gamma^2;1,1) &=& 
\frac{1}{\epsilon} ~+~ 2 ~-~ \ln \left( \frac{C_A \gamma^4}{\mu^4} \right) 
\nonumber \\
&& -~ \frac{1}{\sqrt{2}} \left[ \left[ 
\sqrt{\left(1+\frac{16C_A\gamma^4}{(p^2)^2} \right)} + 1 \right]^{1/2} \left[ 
\frac{1}{2} \ln \left( \frac{16C_A\gamma^4}{(p^2)^2} \right) \right. \right. 
\nonumber \\ 
&& \left. \left. ~~~~~~~~~~-~ \frac{1}{2} \ln 
\left[ 1 + \sqrt{(1+\frac{16C_A\gamma^4}{(p^2)^2}} \right] \right. \right. 
\nonumber \\
&& \left. \left. ~~~~~~~~~~-~ \ln \left[ \left( 1 
+ \sqrt{\left(1+\frac{16C_A\gamma^4}{(p^2)^2} \right)} \right)^{1/2} 
- \sqrt{2} \right] \right] \right. \nonumber \\
&& \left. ~~~+~ \left[ \sqrt{\left(1+\frac{16C_A\gamma^4}{(p^2)^2} \right)} 
- 1 \right]^{1/2} \right. \nonumber \\
&& \left. ~~~~~~~~~\times~ \tan^{-1} \left[ \sqrt{2} \left[ 
\sqrt{\left(1+\frac{16C_A\gamma^4}{(p^2)^2} \right)} - 1 \right]^{-1/2} 
\right] \right. \nonumber \\  
&& \left. ~~~-~ i \left[ \sqrt{\left(1+\frac{16C_A\gamma^4}{(p^2)^2} \right)} 
+ 1 \right]^{1/2} \right. \nonumber \\
&& \left. ~~~~~~~~~\times~ \tan^{-1} \left[ \sqrt{2} \left[ 
\sqrt{\left(1+\frac{16C_A\gamma^4}{(p^2)^2} \right)} - 1 \right]^{-1/2} 
\right] \right. \nonumber \\  
&& \left. ~~~+~ i \left[ \sqrt{\left(1+\frac{16C_A\gamma^4}{(p^2)^2}
\right)} - 1 \right]^{1/2} \left[ \frac{1}{2} \ln \left( 
\frac{16C_A\gamma^4}{(p^2)^2} \right) \right. \right. \nonumber \\ 
&& \left. \left. ~~~~~~-~ \frac{1}{2} \ln 
\left[ 1 + \sqrt{(1+\frac{16C_A\gamma^4}{(p^2)^2}} \right] \right. \right. 
\nonumber \\
&& \left. \left. ~~~~~~-~ \ln \left[ \left( 1 
+ \sqrt{\left(1+\frac{16C_A\gamma^4}{(p^2)^2} \right)} \right)^{1/2} 
- \sqrt{2} \right] \right] \right] ~+~ O(\epsilon) \nonumber \\  
\label{masint1} 
\end{eqnarray}  
and its conjugate, and 
\begin{eqnarray}
I_1(p,i\sqrt{C_A}\gamma^2,-i\sqrt{C_A}\gamma^2;1,1) &=& 
\frac{1}{\epsilon} ~+~ 2 ~-~ \frac{[p^2 - 2 i \sqrt{C_A}\gamma^2]}{2p^2} 
\ln \left( \frac{i \sqrt{C_A} \gamma^2}{\mu^2} \right) \nonumber \\
&& -~ \frac{[p^2 + 2 i \sqrt{C_A}\gamma^2]}{2p^2} 
\ln \left( \frac{ - i \sqrt{C_A} \gamma^2}{\mu^2} \right) \nonumber \\
&& -~ \frac{\sqrt{\left[4 C_A \gamma^4 - (p^2)^2 \right]}}{p^2} \tan^{-1}
\left[ -~ \frac{\sqrt{\left[ 4 C_A \gamma^4 - (p^2)^2 \right]}}{p^2} \right] 
\nonumber \\
&& +~ O(\epsilon) ~. 
\label{masint2} 
\end{eqnarray}  
When one of the masses $m_i^2$ is zero, we use the result
\begin{eqnarray}
\int_k \frac{1}{k^2 [(k-p)^2 + i \sqrt{C_A} \gamma^2]} &=& 
\frac{1}{\epsilon} ~+~ 2 ~-~ \ln \left( \frac{i \sqrt{C_A} \gamma^2}{\mu^2} 
\right) \nonumber \\
&& -~ \frac{[p^2 + i \sqrt{C_A} \gamma^2]}{p^2} \ln \left[ 
\frac{[p^2 + i \sqrt{C_A} \gamma^2]}{i \sqrt{C_A} \gamma^2} \right] ~+~ 
O(\epsilon) 
\end{eqnarray} 
and its conjugate. Finally, for completeness the one loop vacuum bubble is
\begin{equation}
\int_k \frac{1}{[k^2+i\sqrt{C_A}\gamma^2]^\alpha} ~=~ 
\frac{\Gamma(1-\half d)}{(4\pi)^{d/2}} \left( i \sqrt{C_A} \gamma^2 
\right)^{\half d - \alpha} 
\label{masint3} 
\end{equation}
which, together with its conjugate, is expanded in powers of $\epsilon$ and
the resulting logarithms treated with
\begin{equation}
\ln \left( i \sqrt{C_A} \gamma^2 \right) ~=~ \frac{1}{2} \ln \left( 
C_A \gamma^4 \right) ~+~ \frac{i\pi}{2} 
\end{equation} 
and its conjugate. We now discuss the $2$-point functions themselves but first 
note that they have to be renormalized. These infinities are absorbed by the 
wave function renormalization constants given in section 2, and we follow 
the procedure used in \cite{50} to remove them in an automatic calculation. 
Though we note that the mixed $2$-point function is in fact finite as a
consequence of the Slavnov-Taylor identity relating $Z_\gamma$ to $Z_A$ and 
$Z_c$. Given the rather involved form for the master integrals, (\ref{masint1})
and (\ref{masint2}), the full explicit expressions are large and do not serve 
to illustrate any major points which can be accessed by other methods we will 
discuss. Since we are primarily interested in the zero momentum limit, we will 
concentrate on extracting that behaviour. There are two ways of doing this. One
is to expand the explicit expressions in powers of $p^2$ and truncate at the 
appropriate term. A second way is to return to the Feynman diagrams themselves 
and carry out the expansion of each integral in powers of $p^2$. This is known 
as the vacuum bubble expansion and is one method to renormalize a quantum field 
theory. Though of course applying this technique here would require the term 
which is finite in $\epsilon$. We have chosen to do this calculation in 
addition to the exact calculation for two reasons. First, it provides an 
alternative check on the explicit result where the expansion of the exact 
result in powers of $p^2$ must agree with the vacuum bubble expansion. Second, 
if one were to go beyond one loop to carry out a two loop analysis, then the 
{\em exact} evaluation of the two loop integrals with masses in the set $\{0, 
i \sqrt{C_A} \gamma^2, - i \sqrt{C_A} \gamma^2 \}$ would be a formidable task 
even if all the massive master integrals were known in closed form explicitly 
before analytic continuation. The vacuum bubble expansion provides a more 
practical and efficient approach. To proceed with the expansion, the 
propagators of the master integral $I_1(p,m_1^2,m_2^2;\alpha,\beta)$, for 
example, can be expanded in powers of $p^2$ where the $p$-dependent propagator 
is expanded recursively with
\begin{equation}
\frac{1}{[(k-p)^2+m^2]} ~=~ \frac{1}{[k^2+m^2]} ~+~ 
\frac{2kp-p^2}{[k^2+m^2][(k-p)^2+m^2]} 
\label{decomp}
\end{equation}  
which is readily implemented in {\sc Form}. The truncation criterion is that
the $2$-point functions themselves are $O\left( (p^2)^2 \right)$. Given the 
fact that (\ref{laggz}) is renormalizable then the $O\left( (p^2)^2 \right)$
terms will be $\epsilon$-finite. However, in performing this bubble expansion 
we cannot apply the identity (\ref{decomp}) when all the masses in a Feynman
integral are zero. This is because such an integral has its momentum dependence
predetermined by $(p^2)^\Gamma$ where $\Gamma$ is the dimension of the
integral. As the momentum dependence is fixed due to the masslessness of the
graph one needs to first isolate such integrals prior to applying the vacuum
bubble expansion to the remaining terms in the decomposition of the overall  
Feynman graph. Once the vacuum bubble expansion has been performed it is
elementary to replace the vacuum bubbles with the general result 
(\ref{masint3}).

Given this algorithm it is straightforward to implement it in {\sc Form} and we
record the $p^2$ expansion of each of the $2$-point functions. We find
\begin{eqnarray}
\langle A^a_\mu(-p) A^b_\nu(p) \rangle &=& \delta^{ab} \left[ p^2 ~+~ 
\left( \left( \frac{15}{32} \ln \left( \frac{p^2}{\mu^2} \right) 
+ \frac{163}{192} \ln \left( \frac{C_A\gamma^4}{\mu^4} \right)  
- \frac{263}{144} + \frac{57\pi}{64} \frac{\sqrt{C_A}\gamma^2}{p^2} \right) 
C_A \right. \right. \nonumber \\
&& \left. \left. ~~~~~~~~~~~~~~~~~~+~ \left( \frac{20}{9} - \frac{4}{3} 
\ln \left( \frac{p^2}{\mu^2} \right) \right) T_F \Nf \right) p^2 a ~+~ O(a^2) 
\right] \nonumber \\
&& +~ O\left((p^2)^2\right) \nonumber \\ 
\langle A^a_\mu(-p) \bar{\phi}^{bc}_\nu(p) \rangle &=& f^{abc} \left[ 1 ~+~ 
\left( \frac{31\pi}{192} \frac{\sqrt{C_A}p^2}{\gamma^2} \right) a ~+~ O(a^2)
\right] \gamma^2 ~+~ O\left((p^2)^2\right) \nonumber \\ 
\langle \phi^{ab}_\mu(-p) \bar{\phi}^{cd}_\nu(p) \rangle &=& \left[  
\delta^{ac} \delta^{bd} \left[ 1 ~-~ \left( \frac{5}{8} - \frac{3}{8} \ln 
\left( \frac{C_A\gamma^4}{\mu^4} \right) \right) a \right] p^2 \right.
\nonumber \\
&& \left. +~ \frac{3}{64} f^{ace} f^{bde} p^2 a ~+~ \frac{1}{24} f^{abe} 
f^{cde} p^2 a ~+~ \frac{9}{32} d_A^{abcd} \frac{p^2}{C_A} a ~+~ O(a^2) 
\right] \nonumber \\
&& +~ O\left((p^2)^2\right)  
\label{twoptres}
\end{eqnarray}
which determine $X$, $M$, $U$, $Q$, $W$, $R$ and $S$. We note that the 
expansion using (\ref{decomp}) agrees with the expansion of the explicit
expressions on the right side of (\ref{masint1}), (\ref{masint2}) and 
(\ref{masint3}). From (\ref{twoptres}) we can deduce several interesting 
properties of the $p^2$~$\rightarrow$~$0$ limit. First, it is clear that at one
loop the gluon {\em propagator}, (\ref{propform}), vanishes as the momentum 
vanishes. This is because when one substitutes the explicit values for $X$, 
$M$, $U$, $Q$, $W$, $R$ and $S$ from (\ref{twoptres}) into the expression for
$A$ in (\ref{propform}) then one finds
\begin{equation}
A ~=~ \left[ \frac{p^2}{C_A\gamma^4} \left[ \frac{3}{8} \ln \left( 
\frac{C_A\gamma^4}{\mu^4} \right) ~-~ \frac{215}{384} \right] a ~+~ 
O \left( (p^2)^2 \right) \right] ~+~ O(a^2) ~. 
\end{equation}  
The leading terms in this expression derive from only $Q$, $R$ and $W$. This
vanishing of the gluon propagator at one loop is then consistent with general 
expectations and lattice results. See, for instance, \cite{51}. Further, if one
examines the $\delta^{ac} \delta^{bd}$ channel of the $\phi^{ab}_\mu$ 
propagator the one loop mass gap condition (\ref{gap1}) alters the $p^2$ 
dependence in a way similar to what occurs in both the $c^a$ and 
$\omega^{ab}_\mu$ ghost propagators. Moreover, this is the channel which 
appears in the Lagrangian itself and this implies that if there was no 
complication from the mixing then there would be an enhancement in that 
channel.

Finally, we note that the {\em full} explicit expressions for the one loop 
corrections to the $2$-point functions are recorded in appendix A. Whilst these
can be substituted into the formal expressions for the propagators, 
(\ref{propform}), the explicit expressions do not lead to any further insight.
Though the formal functional form may be of use in obtaining better 
parametrizations of lattice and DSE data in the low $p^2$ region. However, for 
the remaining fields which are the Faddeev-Popov ghost and the anticommuting 
Zwanziger ghost, the propagators are  
\begin{eqnarray}  
D_c(p^2) &=& \left[ -~ 1 ~+~ \left[ \frac{5}{4} ~-~ \frac{3}{8} \ln \left( 
\frac{C_A \gamma^4}{\mu^4} \right) ~+~ \frac{3 \sqrt{C_A} \gamma^2}{4p^2} 
\tan^{-1} \left[ \frac{\sqrt{C_A}\gamma^2}{p^2} \right] \right. \right. 
\nonumber \\
&& \left. \left. ~~~~~~~~~~~~~-~ \frac{3 \pi \sqrt{C_A} \gamma^2}{8p^2} ~+~ 
\frac{C_A\gamma^4}{8(p^2)^2} \ln \left[ 1 + \frac{(p^2)^2}{C_A\gamma^4} 
\right] ~-~ \frac{3}{8} \ln \left[ 1 + \frac{(p^2)^2}{C_A\gamma^4} \right]
\right. \right. \nonumber \\
&& \left. \left. ~~~~~~~~~~~~~-~ \frac{p^2}{4\sqrt{C_A}\gamma^2} \tan^{-1} 
\left[ \frac{\sqrt{C_A}\gamma^2}{p^2} \right] \right] C_A a \right]^{-1} ~+~ 
O(a^2) 
\end{eqnarray}  
where the ghost form factor is defined by  
\begin{equation} 
\langle c^a(p) \bar{c}^b(-p) \rangle ~=~ \frac{D_c(p^2)}{p^2} \delta^{ab} ~. 
\end{equation}
Using the gap equation, (\ref{gap1}), we have  
\begin{eqnarray}  
D_c(p^2) &=& \left[ \left[ \frac{5}{8} ~+~ 
\frac{\pi p^2}{8\sqrt{C_A}\gamma^2} ~+~ \frac{3 \sqrt{C_A} \gamma^2}{4p^2} 
\tan^{-1} \left[ \frac{\sqrt{C_A}\gamma^2}{p^2} \right] 
\right. \right. \nonumber \\
&& \left. \left. ~~-~ \frac{3 \pi \sqrt{C_A} \gamma^2}{8p^2} ~+~ 
\frac{C_A\gamma^4}{8(p^2)^2} \ln \left[ 1 + \frac{(p^2)^2}{C_A\gamma^4} 
\right] \right. \right. \nonumber \\
&& \left. \left. ~~-~ \frac{3}{8} \ln \left[ 1 + \frac{(p^2)^2}{C_A\gamma^4} 
\right] ~-~ \frac{p^2}{4\sqrt{C_A}\gamma^2} \tan^{-1} \left[ 
\frac{\sqrt{C_A}\gamma^2}{p^2} \right] \right] C_A a \right]^{-1} ~+~ 
O(a^2) ~. \nonumber \\ 
\end{eqnarray}  
So that $D^{-1}_c(p^2)$ is $O((p^2)^2)$ as $p^2$ $\rightarrow$ $0$. For 
$\omega^{ab}_\mu$ it turns out that defining 
\begin{equation}
\langle \omega^{ab}_\mu(p) \bar{\omega}^{cd}_\nu(-p) \rangle ~=~ 
\delta^{ac}\delta^{bd}\frac{D_\omega(p^2)}{p^2} \eta_{\mu\nu} 
\end{equation}
then
\begin{equation}
D_\omega(p^2) ~=~ D_c(p^2)
\label{omceq} 
\end{equation}
to one loop and no longitudinal component is generated. So the 
$\omega^{ab}_\mu$ ghost enhancement actually follows trivially from the 
equivalence with the Faddeev-Popov ghost for all momenta. This follows as a 
result of our earlier observation of the factoring of the Lorentz structure 
from the $2$-point function. Given there is a similar enhancement at two loops 
from the vacuum bubble expansion, it is tempting to speculate that not only 
does (\ref{omceq}) hold at two loops but maybe also to all orders. Whilst we 
have noted that there is a similar type of enhancement of $Q$ in the 
$\phi^{ab}_\mu$-propagator, it is also turns out that 
$Q$~$=$~$-$~$\left(D_c(p^2)\right)^{-1}$~$=$~$-$~$\left(D_\omega(p^2)\right)^{-1}$ 
at one loop. Finally for the quark propagator, where the quarks are massless, 
defining the form factor by 
\begin{equation}
\langle \psi^{iI}(p) \bar{\psi}^{jJ}(-p) \rangle ~=~ i \delta^{ij} \delta^{IJ} 
D_\psi(p^2) \frac{\pslash}{p^2} 
\end{equation}
then
\begin{eqnarray}
D_\psi(p^2) &=& \left[ 1 ~+~ \left[ \frac{1}{2} ~-~ 
\frac{C_A\gamma^4}{2(p^2)^2} \ln \left[ 1 + \frac{(p^2)^2}{C_A\gamma^4} 
\right] ~-~ \frac{3\sqrt{C_A}\gamma^2}{p^2} \tan^{-1} \left[ 
\frac{\sqrt{C_A}\gamma^2}{p^2} \right] \right. \right. \nonumber \\
&& \left. \left. ~~~~~~~~~+~ \frac{3\pi}{4} \frac{\sqrt{C_A}\gamma^2}{p^2} ~-~ 
\frac{p^2}{2\sqrt{C_A}\gamma^2} \tan^{-1} \left[ \frac{\sqrt{C_A}\gamma^2}{p^2}
\right] \right] C_F a \right]^{-1} \nonumber \\
&& +~ O(a^2) ~.
\end{eqnarray}  
We note that in the limit $p^2$ $\rightarrow$ $0$ then
\begin{equation}
D_\psi(p^2) ~=~ 1 ~+~ \left[ \frac{3}{2} ~-~ 
\frac{\pi p^2}{4\sqrt{C_A}\gamma^2} ~+~ O(p^2) \right] C_F a ~+~ O(a^2)
\end{equation}
so that the form factor tends to a constant at zero momentum. 
 
\sect{$\alpha_S$ freeze-out.}

Having examined the zero momentum limit of the $2$-point functions and form
factors, we now apply the results to the problem of the value of the effective 
strong coupling constant in the same limit. In both lattice and DSE analyses 
one can study the value of a renormalization group invariant effective coupling
constant, denoted by $\alpha^{\mbox{\footnotesize{eff}}}_S (p^2)$, which is 
defined as a result of the renormalization properties of the gluon ghost vertex
in the Landau gauge. In terms of the gluon and Faddeev-Popov form factors, 
$D_A(p^2)$ and $D_c(p^2)$, where  
\begin{equation}
\langle A^a_\mu(p) A^b_\nu(-p) \rangle ~=~ \delta^{ab} \frac{D_A(p^2)}{p^2} 
P_{\mu\nu}(p) 
\end{equation}
the effective coupling is 
\begin{equation}
\alpha^{\mbox{\footnotesize{eff}}}_S (p^2) ~=~ \alpha(\mu) D_A(p^2) \left(
D_c(p^2) \right)^2 
\label{effccdef} 
\end{equation}
where $\alpha(\mu)$ is the running strong coupling constant and  
$\alpha$~$=$~$g^2/(4\pi)$. Due to the gluon ghost Slavnov-Taylor identity,
\cite{52}, there is no contribution from the vertex form factor. Thus, given
that we have one loop expressions for $D_A(p^2)$ and $D_c(p^2)$ in the  
$p^2$~$\rightarrow$~$0$ limit, it is straightforward to examine the structure
of (\ref{effccdef}) in the same limit. 

First, if we consider the expression with the tree level form factors
\begin{equation}
D_A(p^2) ~=~ \frac{(p^2)^2}{[(p^2)^2+C_A\gamma^4]} ~+~ O(a) ~~~,~~~ 
D_c(p^2) ~=~ 1 ~+~ O(a) 
\end{equation} 
then clearly 
\begin{equation}
\alpha^{\mbox{\footnotesize{eff}}}_S (p^2) ~=~ \frac{\alpha(\mu) (p^2)^2}
{[(p^2)^2+C_A\gamma^4]} 
\end{equation}
which vanishes as $p^2$~$\rightarrow$~$0$. However, if one includes the one
loop corrections then with 
\begin{equation}
D_c(p^2) ~=~ \left[ -~ 1 ~+~ \left[ \frac{5}{8} - \frac{3}{8} \ln \left( 
\frac{C_A \gamma^4}{\mu^4} \right) ~-~ \left( \frac{\pi}{8} 
\frac{p^2}{\sqrt{C_A}\gamma^2} \right) ~+~ O(p^4) \right] C_A a ~+~ O(a^2) 
\right]^{-1} 
\end{equation}  
where we have expanded the propagator to the next term in the momentum
expansion, we have 
\begin{equation} 
\alpha^{\mbox{\footnotesize{eff}}}_S (0) ~=~ \lim_{p^2 \rightarrow 0} \left[  
\frac{ \alpha(\mu) \left[ 1 + C_A \left( \frac{3}{8} \ln \left( 
\frac{C_A \gamma^4(\mu)}{\mu^4} \right) - \frac{215}{384} \right) a(\mu) 
\right] (p^2)^2 } 
{ C_A \gamma^4(\mu) \left[ 1 + C_A \left( \frac{3}{8} \ln \left( 
\frac{C_A \gamma^4(\mu)}{\mu^4} \right) - \frac{5}{8} 
+ \frac{\pi p^2}{8 \sqrt{C_A} \gamma^2(\mu)} \right) a(\mu) \right]^2 } 
\right] ~. 
\end{equation}  
In this expression we have expanded the denominator factor of $D_A(p^2)$ and 
included it in the $O(p^2)$ part of the numerator. Though in fact it will not 
contribute to the final value of $\alpha^{\mbox{\footnotesize{eff}}}_S (0)$ 
which is derived from the displayed numerator and denominator factors. Before 
taking the limit one can use Gribov's gap equation, (\ref{gap1}), to enforce 
the ghost enhancement in the ghost form factor which leaves the denominator as 
$O(a^2)$. This coupling constant dependence in fact cancels with similar 
factors in the numerator. One comes from the $\alpha(\mu)$ in the definition 
and another one derives from the numerator factor when the gap equation is 
used. Unlike the cancellation in the ghost propagator to give ghost 
enhancement, in the numerator the numerical term at one loop does not match 
that of the gap equation and one is left with an $O((p^2)^2)$ term in the 
numerator. This momentum dependence then cancels the $(p^2)^2$ term on the 
denominator. Finally, the running Gribov mass parameter, $\gamma(\mu)$, also 
cancels with one factor of $\gamma^4$ deriving from the determinant of the 
$2$-point functions and one from the $p^2/\gamma^2$ term of the expansion of 
the Faddeev-Popov ghost $2$-point function. Since the running coupling constant
and dimensionful quantities also cancel one is left with a pure number when 
$p^2$~$\rightarrow$~$0$. We find 
\begin{equation} 
\alpha^{\mbox{\footnotesize{eff}}}_S (0) ~=~ \frac{50}{3\pi C_A} ~. 
\label{alfre}
\end{equation}  
Specifying various gauge groups of interest, we have 
\begin{equation} 
\left. \alpha^{\mbox{\footnotesize{eff}}}_S (0) \right|_{SU(3)} ~=~ 1.7684
\label{fresu3} 
\end{equation}  
and 
\begin{equation} 
\left. \alpha^{\mbox{\footnotesize{eff}}}_S (0) \right|_{SU(2)} ~=~ 2.6526 ~.
\end{equation}  
This is a novel property of the Gribov-Zwanziger Lagrangian. From studying the
calculation in the way we have presented it, we note that the gap equation was
essential in removing the tree terms of the form factors, as well as providing
factors of $a(\mu)$ and $\gamma(\mu)$ which {\em precisely} cancel. Indeed the
role of $\gamma(\mu)$ seems crucial since it provides the balancing
dimensionality for the momentum which eventually vanishes. More significantly 
it is interesting to note that the cancellation between the numerator and 
denominator factors is between pieces which involve the two types of ghost. In 
other words the anticommuting Faddeev-Popov ghost and the commuting Zwanziger 
ghost, $\phi^{ab}_\mu$. The contribution from the former, which is required for
the usual gauge fixing, is clearly demanded by the definition but the latter 
arises through the implementation of the Gribov horizon which is, of course, an
infrared contribution originating from the ambiguity of the gauge fixing 
procedure.

Concerning the actual numerical value of 
$\alpha^{\mbox{\footnotesize{eff}}}_S (0)$ in $SU(3)$, (\ref{fresu3}), it is 
worth contrasting with estimates from other methods. From a survey of articles 
(which is by no means exhaustive) it appears that the non-zero values fall into
three classes. These are when $\alpha^{\mbox{\footnotesize{eff}}}_S (0)$ is 
deduced from experimental data (P), from numerical studies such as lattice or 
DSE and from more analytic approaches, (A). Currently there is not a consensus 
of values from these three divisions which are summarised in Table 1 where the 
result of \cite{18} is the central value of the range the authors quote. For 
instance, the DSE value differs from the phenomenology based approach by an 
order of magnitude; with analytic approaches, including this article, bridging 
between the two extremes. However, in this context we note that (\ref{alfre}) 
should be viewed as a qualitative result rather than quantitative. First, the 
computation, whilst simple in its derivation, is only at one loop. Clearly
there are higher order corrections to the form factors. Indeed it is not
immediately apparent how the $O(a^2)$ corrections would conspire to cancel to
leave a result independent of $a(\mu)$ and $\gamma(\mu)$ as was the case at
leading order. However, it is also not clear whether comparison of  
$\alpha^{\mbox{\footnotesize{eff}}}_S (0)$ from different methods has any 
meaning in the first instance since it corresponds to the evaluation of a
quantity defined in perturbation theory but evaluated in the infrared limit.
Though the lattice and DSE estimates are for the same quantity as 
(\ref{effccdef}) in the Landau gauge. Moreover, (\ref{effccdef}) depends on the
ghost form factor which would not be an object immediately accessible 
experimentally. In this context it is worth commenting on a recent proposal of 
\cite{32,33} where an effective coupling constant is defined from other 
vertices in the Gribov-Zwanziger Lagrangian. For instance, it was argued that 
one could examine the form factor of the triple gluon vertex, $D_{AAA}(p^2)$, 
and derive an effective coupling with $D_A(p^2)$, defined with respect to some
momentum configuration for the external gluons, with 
\begin{equation}
\alpha^{\mbox{\footnotesize{eff}}}_{S \, AAA} (p^2) ~=~ \alpha(\mu)
D_{AAA}(p^2) \left( D_A(p^2) \right)^3 
\end{equation}
where $D_{AAA}(p^2)$ is necessary due to the lack of a Slavnov-Taylor identity 
comparable to that of the gluon ghost vertex. To have a freezing of 
$\alpha^{\mbox{\footnotesize{eff}}}_{S \, AAA} (p^2)$, given the behaviour of 
$D_A(p^2)$ established here, then $D_{AAA}(p^2)$ would have to diverge in such 
a way that the momentum dependence cancels in the zero momentum limit,
\cite{32,33}. Whilst this is essentially a dimensional observation, 
\cite{32,33}, it does not necessarily follow that the {\em same} numerical 
value for the frozen effective coupling would emerge. Given that we now have a
renormalizable localized Lagrangian implementing the Gribov horizon, one
natural question which could be considered is whether this behaviour for the
triple gluon vertex could be determined in the $p^2$~$\rightarrow$~$0$ limit. 
Although such a calculation is beyond the scope of the current article, we note
that on renormalizability grounds $D_{AAA}(p^2)$ would have to be finite as 
$p^2$~$\rightarrow$~$0$. Therefore, if one is to extract a value for  
$\alpha^{\mbox{\footnotesize{eff}}}_{S \, AAA} (0)$, via the suggestion of
\cite{32,33}, it would seem to us that the behaviour is truly driven by some
non-perturbative mechanism.  
{\begin{table}[ht]  
\begin{center} 
\begin{tabular}{|c||c|c|} 
\hline 
$\alpha^{\mbox{\footnotesize{eff}}}_S (0)$ & Method & Reference \\ 
\hline 
 0.47 & AP & \cite{20,21} \\ 
 0.56 & P & \cite{18} \\ 
 0.60 & P & \cite{24,25,26} \\ 
 0.60 & P & \cite{19} \\ 
 0.63 & P & \cite{22,23} \\ 
 0.82 & P & \cite{27,28} \\ 
\hline 
 1.40 & A & \cite{29,30,31} \\ 
 1.77 & A & this study \\ 
\hline 
 2.97 & DSE & \cite{32,33} \\ 
\hline 
\end{tabular} 
\end{center} 
\begin{center} 
{Table 1. Non-zero estimates of $\alpha^{\mbox{\footnotesize{eff}}}_S (0)$.} 
\end{center} 
\end{table}}  

Although we have given a summary of the non-zero finite values for 
$\alpha^{\mbox{\footnotesize{eff}}}_S (0)$ it should be noted that various
lattice and DSE studies find a value of zero as $p^2$~$\rightarrow$~$0$,
(see, for example, \cite{34,37}), for the effective coupling defined from the
gluon and ghost form factors. Further, lattice and DSE studies of effective 
couplings defined from other vertices also find a zero value in this limit. For
instance, see \cite{35,36}. Therefore, it would appear that currently there is 
no common view of the precise zero momentum behaviour. However, these latter 
studies are of the effective coupling defined from the triple gluon and quark 
gluon vertices respectively. So if our argument about the finiteness of
$D_{AAA}(p^2)$ as $p^2$~$\rightarrow$~$0$ from (\ref{laggz}) is valid, then the
vanishing of $D_A(0)$ would imply that 
$\alpha^{\mbox{\footnotesize{eff}}}_{S \, AAA} (0)$~$=$~$0$ which would appear 
to be consistent with the analyses of this vertex, \cite{35}. 

The situation for $SU(2)$ is similar, though of course the only results 
available are from lattice and DSE studies. Briefly, in the same way that 
certain $SU(3)$ lattice and DSE computations give similar non-zero values for 
$\alpha^{\mbox{\footnotesize{eff}}}_S (0)$ the same is true for $SU(2)$ with
the common freeze-out value of 
$\alpha^{\mbox{\footnotesize{eff}}}_S (0)$~$=$~$5(1)$, \cite{53}, or $5.2$,
\cite{54}. Whilst this is significantly larger than our value of $2.65$ there 
is at least the consistent observation that our $SU(2)$ value is larger than 
the $SU(3)$ value. Though from the explicit expression this is primarily due to
the denominator color factor in (\ref{alfre}). Taking the ratio of the $SU(2)$ 
and $SU(3)$ values for $\alpha^{\mbox{\footnotesize{eff}}}_S (0)$ from the DSE 
analyses one finds the ratio $1.75$ in comparison with the ratio of $1.5$ of 
(\ref{alfre}).  

Finally, in comparing (\ref{alfre}) with DSE estimates it is worth noting that
the latter are invariably computed in MOM schemes which are mass dependent 
renormalization schemes. By contrast (\ref{alfre}) has been deduced in the 
$\MSbar$ scheme which is a mass independent scheme. Whilst ultimately the 
definition of $\alpha^{\mbox{\footnotesize{eff}}}_S (p^2)$ is a renormalization
group invariant, if the quantity (\ref{effccdef}) has a meaning in the 
infrared then the full value will be scheme independent. Therefore, computing 
the two loop corrections to (\ref{alfre}) would be useful to see if there is 
convergence to a higher value. In discussing some of the issues concerning the 
freezing from a theoretical point of view, it is worth noting the one loop 
study of \cite{55} where the infrared behaviour of an effective coupling 
defined via the quark gluon vertex with massive quarks in an arbitrary
covariant gauge in the MOM scheme. There the presence of the gauge parameter 
led to differing zero momentum behaviours indicative of the subtleties 
associated with analysing the infrared behaviour of a quantity which has only 
justifiable meaning in the ultraviolet r\'{e}gime. 

\sect{Power corrections.} 

Having studied the effective coupling in the limit of zero momentum, using both
the bubble expansion and the exact expressions for the $2$-point functions, we
can now examine the propagators and the couplings in another limit. In 
\cite{43,44}, a numerical fit of the lattice computation of the effective 
coupling constant suggested that in a certain momentum range, the coupling 
deviated from the expected (perturbative) behaviour by a piece which could be 
parametrized by power corrections. Intriguingly these corrections were of the
form $O(1/p^2)$ and not the expected $O(1/(p^2)^2)$ behaviour. In other words,
\cite{43,44}, 
\begin{equation}
\alpha^{\mbox{\footnotesize{eff}}}_S (p^2) ~=~ 
\alpha^{\mbox{\footnotesize{pert}}}_S (p^2) ~+~ \frac{c_2}{p^2} ~+~ 
O \left( \frac{1}{(p^2)^2} \right) ~. 
\end{equation} 
An $O(1/(p^2)^2)$ correction is motivated by the fact that to match the 
dimensionality of the momentum dependence one needs a dimension four quantity. 
Given the fact that the operator product expansion and sum rule studies of
\cite{56} suggest the condensation of the gauge invariant operator 
$G^a_{\mu\nu} G^{a\,\mu\nu}$, then there is an a priori clear candidate for
the numerator which is $\langle G^a_{\mu\nu} G^{a\,\mu\nu} \rangle$. However,
in order to have an $O(1/p^2)$ correction, one would have to have a lower 
dimensional operator condensing. Since one such operator is $\half A^a_\mu
A^{a\,\mu}$, this has led to the suggestion that this operator condenses,
\cite{43}, and consequently estimates for $\langle \half A^a_\mu A^{a\,\mu} 
\rangle$ have been derived from the operator product expansion, \cite{57},
and the local composite operator method, \cite{58}. Indeed it was observed in
\cite{59,60} that the perturbative QCD vacuum was unstable and the vacuum
expectation value of such operators, and its non-local gauge invariant
generalization, would play a significant role in understanding the true vacuum. 
Also, the condensation of such a dimension two operator was already discussed 
in the Coulomb gauge in \cite{61}. Clearly $\half A^a_\mu A^{a\,\mu}$ is a 
gauge variant operator but the appearance of its vacuum expectation value in 
the fits of \cite{43,44} is neither unexpected nor inconsistent with this since
the effective coupling constant is gauge dependent in mass dependent 
renormalization schemes such as the MOM scheme used in \cite{43,44}. As a 
consequence of these observations concerning the apparent existence of a 
non-zero value for $\langle \half A^a_\mu A^{a\,\mu} \rangle$ there has been 
renewed interest in trying to understand the dynamical generation of a gluon 
mass. However, given that the Gribov mass, $\gamma^2$, can effectively 
reproduce a gluon mass, it is worth investigating whether one can {\em mimic} 
the power correction behaviour of \cite{43,44} from the Gribov-Zwanziger 
Lagrangian, (\ref{laggz}). Therefore, we have examined the integrals used to 
construct the form factor corrections exactly and expanded them in powers of 
$\sqrt{C_A}\gamma^2$. Though we remark that in the expansion in which we work 
$\sqrt{C_A} \gamma^2$~$<$~$p^2$, so that we need to rewrite one of the parts of
an exact integral through the replacement 
\begin{eqnarray}
&& \sqrt{\left[ 4 C_A \gamma^4 - (p^2)^2 \right]} \tan^{-1} \left[ -~
\frac{\sqrt{\left[ 4 C_A \gamma^4 - (p^2)^2 \right]}}{p^2} \right] \nonumber \\
&& \mapsto ~ \sqrt{\left[ (p^2)^2 - 4 C_A \gamma^4 \right]} \ln \left[ 
\frac{\left[ p^2 + \sqrt{\left[ (p^2)^2 - 4 C_A \gamma^4 \right]}\right]} 
{\left[ p^2 - \sqrt{\left[ (p^2)^2 - 4 C_A \gamma^4 \right]}\right]} \right] ~.
\end{eqnarray} 
Therefore expanding the $2$-point functions in powers of $\gamma^2$, produces
\begin{eqnarray}
X &=& \left[ \left[ \left( \frac{13}{6} \ln \left( \frac{p^2}{\mu} \right) 
- \frac{97}{36} \right) p^2 + \frac{3\pi\sqrt{C_A}\gamma^2}{8} \right] C_A 
\right. \nonumber \\
&& \left. ~-~ \left[ \frac{4}{3} \ln \left( \frac{p^2}{\mu} \right) 
- \frac{20}{9} \right] T_F \Nf p^2 ~+~ O(\gamma^4) \right] a ~+~ O(a^2)
\nonumber \\
M &=& U ~=~ \left[ \frac{11C_A}{8} \gamma^2 ~+~ O(\gamma^4) \right] a ~+~ 
O(a^2) \nonumber \\ 
Q &=& \left[ \left[ ~-~ \left( 1 - \frac{3}{4} \ln \left( \frac{p^2}{\mu} 
\right) \right) p^2 ~+~ \frac{3\pi\sqrt{C_A}\gamma^2}{8} \right] C_A ~+~ 
O(\gamma^4) \right] a ~+~ O(a^2) \nonumber \\
W &=& R ~=~ S ~=~ O(a^2)   
\end{eqnarray} 
and 
\begin{equation}
D_c(p^2) ~=~ \left[ -~ 1 ~+~ \left[ 1 - \frac{3}{4} \ln \left( 
\frac{p^2}{\mu^2} \right) ~-~ \frac{3\pi}{8} 
\frac{\sqrt{C_A}\gamma^2}{p^2} ~+~ O(\gamma^4) \right] C_A a ~+~ O(a^2) 
\right]^{-1} 
\end{equation}  
where we have neglected the term beyond the first $\gamma^2/p^2$ correction in
each case. For the gluon and Faddeev-Popov ghost the finite 
$\gamma$-independent pieces both agree with the result of the massless
calculation.

Interestingly the gluon propagator correction is $O(\gamma^2)$ and not 
$O(\gamma^4)$ as might have been suggested from the original gluon propagator
which is a function of $\gamma^4$. Although the correction is from the one 
loop term and involves $a$, it qualitatively introduces an effective mass for
the gluon in this next to high energy limit. One comment concerning the choice
of the sign of $\gamma^2$ is worth making. Eliminating the auxiliary scalar 
field $\phi^{ab}_\mu$ from (\ref{laggz}) means that (\ref{laggz}) is an even 
function of $\gamma^2$. One could fix the sign of $\gamma^2$ by arguing that it
has to produce a non-tachyonic effective propagator of the usual form from the 
binomial expansion. However, one could equally choose the mass to be tachyonic 
without upsetting the two loop gap equation and the ghost enhancement. Although
this may not appear to be a physically sensible choice, we remark that the 
study of \cite{62} suggested that introducing an effective tachyonic gluon mass
into current correlators could reproduce experimental data more accurately 
compared to the case of no gluon mass. Whilst there did not appear to be any 
justification for such a tachyonic gluon in \cite{62}, the freedom of choosing 
the sign of $\gamma^2$, whilst still {\em retaining} the infrared properties 
already discussed, does appear appealing. 

In addition we can now examine the effective coupling constant,
(\ref{effccdef}), in the same limit. Retaining only the $O(\gamma^2)$ piece
of $A$ from the $2$-point functions, then
\begin{equation}
D_A(p^2) ~=~ 1 ~-~ \frac{3C_A\pi}{8} \frac{\sqrt{C_A}\gamma^2}{p^2} a ~+~ 
O \left( \frac{1}{(p^2)^2} \right) ~.  
\end{equation}
In expressing the power corrections in this way, we are treating it as an 
expansion in powers of $\sqrt{C_A}\gamma^2/p^2$ since the appropriate factor of
$C_A$ is associated with the Gribov mass. Thus from the definition, and 
regarding the $\gamma$ independent piece as corresponding to 
$\alpha^{\mbox{\footnotesize{pert}}}_S (p^2)$, then to one
loop we have 
\begin{equation}
\alpha^{\mbox{\footnotesize{eff}}}_S (p^2) ~=~ 
\alpha^{\mbox{\footnotesize{pert}}}_S (p^2) \left[ 1 ~-~ \frac{9C_A\pi}{8} 
\frac{\sqrt{C_A}\gamma^2}{p^2} a ~+~ O \left( \frac{1}{(p^2)^2} \right) \right] 
\label{ccpower}
\end{equation} 
in $\MSbar$. From the $O(\gamma^2)$ corrections one {\em formally} produces an 
$O(1/p^2)$ term as the leading power correction as opposed to an $O(1/(p^2)^2)$
one. Also the coefficient has the opposite sign to that of \cite{43,44} and 
would be in keeping with our suggested choice of changing the sign of 
$\gamma^2$. However, since that calculation was in the MOM scheme in contrast 
to the $\MSbar$ scheme {\em one} loop computation here, this would not
necessarily be sufficient justification for altering the sign of $\gamma^2$.
Again we merely regard it as an interesting observation. Further, we emphasise 
that, as in earlier sections, these are qualitative observations of this 
effective coupling constant deriving from a one loop calculation with the 
motivation being to appreciate the implications of the Gribov parameter in 
comparison with non-perturbative numerical analyses. Clearly such a power 
correction could equally well be produced by the presence of $\langle \half 
A^a_\mu A^{a\,\mu} \rangle$ instead of the condensate (\ref{gapdef}). Indeed 
there have been more recent studies, \cite{39,63,64}, of the effect a non-zero 
$\langle \half A^a_\mu A^{a\,\mu} \rangle$ has in the Gribov-Zwanziger 
Lagrangian in the Landau gauge. We also remark that from the renormalization 
structure, \cite{38,39}, both $\half A^a_\mu A^{a\,\mu}$ and $\gamma^2$ have 
the same anomalous dimensions. So that a combination of both these dimension 
two operator vacuum expectation values could be responsible for the $O(1/p^2)$ 
correction in the data of \cite{43,44}. At a more formal level it is worth
commenting on the structure of the $O(\gamma^2)$ correction in (\ref{ccpower}).
This effective coupling constant is by construction a renormalization group
invariant quantity due to the underlying Slavnov-Taylor identity for this
vertex, \cite{65}. For the Gribov-Zwanziger Lagrangian this identity remains 
valid which can be observed, for instance, from explicit loop calculations of 
the relevant anomalous dimensions using (\ref{laggz}) which reflect the 
ultraviolet position. As is evident from the freezing calculation, the 
numerical value of $\alpha^{\mbox{\footnotesize{eff}}}_S (0)$ is clearly a
renormalization group invariant. However, it is not immediately apparent if the 
power correction, which is manifest in the remaining energy range, also retains 
this feature. To fully examine this situation is not as straightforward as it 
would simply appear. This is primarily due to the fact that the full
renormalization group for (\ref{laggz}) has not yet been constructed. 
Ordinarily this involves parameters and quantities which are {\em independent}.
By contrast, in (\ref{laggz}) the Gribov mass, being a parameter of the theory
which appears in the (ultraviolet) renormalization group equation, is not
independent by virtue of the Gribov gap equation. Therefore, to fully access 
the renormalization group properties of quantities deduced from (\ref{laggz}) 
it would seem necessary in the first instance to develop the renormalization 
group for (\ref{laggz}) subject to the Gribov gap equation constraint. This is 
beyond the scope of this article.  

If we also consider the quark propagator, then in the same limit we find
\begin{equation}
D_\psi(p^2) ~=~ 1 ~+~ \left[ \frac{3\pi \sqrt{C_A} \gamma^2}{4p^2} ~+~
O \left( \frac{1}{(p^2)^2} \right) \right] C_F a ~+~ O(a^2) ~.
\end{equation}  
By the same token that we argued that the power corrections appeared to
generate a mass for the gluons one might regard this correction as a generated
quark mass. Although the sign is opposite to what one observes in the gluon
case but with magnitude $3\pi C_F a/4$, we emphasise that this is in the 
situation where there is no initial quark mass and therefore it does not 
correspond to the part of a fermionic propagator which ordinarily defines the 
mass term.  

Finally, if one accepts that there can be an effect from the Gribov parameter 
in studying power corrections in gauge variant quantities, then considering 
higher power corrections leads to the conclusion that one could have 
contributions from not only $\langle G^a_{\mu\nu} G^{a\,\mu\nu} \rangle$ but 
also $\gamma^4$ in the $O(1/(p^2)^2)$ correction. Therefore, in extracting 
estimates for $\langle G^a_{\mu\nu} G^{a\,\mu\nu} \rangle$ in such 
calculations, one would in principle need to make allowance for the potential 
presence of additional $\gamma^4$ type terms. For the computation of gauge 
invariant quantities, there ought not to be any contribution from the Gribov 
mass. 

\sect{Discussion.} 

In examining the one loop corrections to the propagators in the 
Gribov-Zwanziger Lagrangian several interesting features have emerged. First,
we have shown that the gluon propagator vanishes in the $p^2$~$\rightarrow$~$0$
limit which is consistent with DSE and some lattice studies. More significantly
we have seen to what extent the gap equation satisfied by the Gribov parameter
underlies the infrared structure of the propagators and effective coupling
constant in (\ref{laggz}). In \cite{1} it was used to ensure that the ghost 
propagator was enhanced in the infrared. However, here it has been used to show
that the $\omega^{ab}_\mu$ ghost propagator is also enhanced at two loops, and 
that the behaviour of the tree channel of the $\phi^{ab}_\mu$ propagator is 
also altered in the infrared at one loop. More significantly the gap equation 
played a central role in the freezing of the renormalization group invariant 
effective coupling defined via the gluon Faddeev-Popov ghost vertex and has 
given some insight into how a non-zero finite value emerges. For instance, in 
the definition (\ref{effccdef}) if the ghost propagator does not satisfy the 
gap equation then the factors of $p^2$ in the zero momentum limit will not 
match between the numerator and denominator to leave a momentum independent 
quantity. Any remaining momentum dependence would dominate the 
$p^2$~$\rightarrow$~$0$ behaviour. Since there appears to be a discrepancy in 
various numerical approaches between a non-zero and zero value for 
$\alpha^{\mbox{\footnotesize{eff}}}_S (0)$, it may be due to the Kugo-Ojima 
condition not being fully and {\em precisely} satisfied. Although our one loop 
calculation has led to a freeze-out value which is different from that from DSE
and certain lattice studies, and is regarded as qualitative in the sense 
described earlier, it would be interesting to see what effect the two loop 
corrections have on the numerical estimate. At one loop the result is clearly 
independent of the number of quarks but quark loops will be present at two 
loops. Recently in \cite{66} a comparison was made between a calculation of 
(\ref{effccdef}) in quenched QCD and one where dynamical fermions were used. 
Although both computations suggested 
$\alpha^{\mbox{\footnotesize{eff}}}_S (0)$~$=$~$0$, the peak of the plot of 
$\alpha^{\mbox{\footnotesize{eff}}}_S (p^2)$ is lower in the dynamical case.
Though we note that intriguingly the peak in the quenched case appears to be
around $1.7$ which is similar to (\ref{fresu3}). In concluding this aspect of 
our analysis it would be fair to make the general observation that the full 
resolution as to whether the quantity (\ref{effccdef}) freezes at a zero or 
non-zero value in the infrared has not yet been achieved and our value is 
merely another contribution to that debate. 
 
Another aspect of this study has been to appreciate the role played by the 
extra Zwanziger ghosts. There are various ways of viewing them. In one both 
$\phi^{ab}_\mu$ and $\omega^{ab}_\mu$ can regarded as being on the same footing
from the way they were originally introduced to localise the Gribov problem. 
However, from the equation of motion for $\phi^{ab}_\mu$, which was used to 
implement the horizon condition, $\phi^{ab}_\mu$ is clearly not unrelated to 
the original gluon field itself and in some sense could be regarded as part of 
a more general spin-$1$ field such as
\begin{equation}
\bar{A}^a_\mu ~=~ A^a_\mu ~+~ \lambda \gamma^2 \frac{1}{\partial^\nu D_\nu} 
A^{a \, \mu} 
\end{equation}
where $\lambda$ is a constant, and which could be regarded as taking account of
some of the more global aspects of the gauge fixing. Indeed the freezing of the
effective coupling was dependent on the value of the $\phi^{ab}_\mu$ $2$-point 
function. So in some respects the infrared behaviour is guided by the inherent 
non-locality of a more general spin-$1$ field. Whilst this field 
$\bar{A}^a_\mu$ is clearly non-local it is similar to the second term in the 
expansion of the gauge invariant spin-$1$ field of \cite{67} which is also 
non-local. In \cite{67} the non-local structure resulting from the gauge 
invariant mass term in the Lagrangian was used to construct a vortex solution 
which underpinned the confinement process. Given that there is now numerical 
evidence for vortices present on the Gribov boundary (in the Coulomb gauge 
\cite{68}), it might be possible to {\em construct} an explicit vortex solution
in the Gribov-Zwanziger Lagrangian and study their dynamics in relation to the 
role they have in confinement. In contrast to the non-local mass term of 
\cite{67}, which is non-localizable since it would require an infinite number 
of extra fields, the non-locality of the Gribov formulation of non-abelian 
gauge fixing is localizable since only the finite set of fields
$\{\phi^{ab}_\mu,\bar{\phi}^{ab}_\mu\}$ and
$\{\omega^{ab}_\mu,\bar{\omega}^{ab}_\mu\}$ are required. A heuristic way of
viewing this is to formally eliminate $\phi^{ab}_\mu$ from (\ref{laggz}), and 
ignoring all anticommuting ghost fields for the moment, then one has   
\begin{equation} 
L^{\mbox{\footnotesize{GZ}}} ~=~ \frac{1}{4} G_{\mu\nu}^a 
G^{a \, \mu\nu} ~-~ \frac{C_A\gamma^4}{2} A^a_\mu \, \frac{1}{\partial^\nu 
D_\nu} A^{a \, \mu} ~+~ \frac{d \NA \gamma^4}{2g^2} 
\label{lagnl} 
\end{equation} 
which involves only one non-local term. In momentum space the quadratic part
of (\ref{lagnl}) clearly leads to the Gribov gluon propagator. Returning to 
$\bar{A}^a_\mu$, one can now view the anticommuting Faddeev-Popov and Zwanziger
ghosts from a different point of view to their original role, if one ignores 
their spins. In other words it seems that one can associate a ghost field with 
each term of $\bar{A}^a_\mu$ and, as we have shown here, each of these ghosts 
have similar infrared properties. For instance, both their propagators and
form factors are equal at one loop and enhance at two loops in the 
$p^2$~$\rightarrow$~$0$ limit. Also the renormalization group invariant 
couplings defined from their interactions with the gluons both freeze at zero 
momentum to the same value. In this respect, aside from their differing spins 
they could be regarded as the first few terms of a ghost multiplet where the 
subsequent terms would be associated with higher order terms of $\bar{A}^a_\mu$
which would involve higher dimension non-local operators. Such operators might
be the ones necessary to give the potential freezing in the effective coupling
constants derived from the triple and quartic gluon vertices as suggested in
\cite{32,33} or the restriction to the fundamental modular region. 

Finally, we note that we have concentrated throughout in this article on 
studying the Gribov-Zwanziger Lagrangian in the Landau gauge only. However, it 
would be interesting to repeat the present analysis for other covariant gauges 
such as the other linear covariant Lorentz gauges and the maximal abelian 
gauge. Indeed for the former gauges there has been preliminary work in this 
direction, \cite{69,70}, and it would be interesting to see, for instance, 
whether the power correction behaviour changed significantly and if the 
variation in magnitude of the residue of the $O(1/p^2)$ correction, if any, 
could be measured, say, on the lattice in a variety of gauges. Moreover, in the
maximal abelian gauge it is possible to define an effective coupling constant
from the Slavnov-Taylor structure of one of the gluon ghost vertices, 
\cite{70}. Therefore, if one obtained the same freeze-out value for this 
effective coupling, it could be evidence for the abelian dominance hypothesis. 
We close by remarking that our study implies that the Gribov-Zwanziger 
Lagrangian has established infrared features which are generally not 
inconsistent with other methods. Specifically, these involve quantities such as 
$\alpha^{\mbox{\footnotesize{eff}}}_S (p^2)$ which do not ultimately involve
the running parameters $g(\mu)$ and $\gamma(\mu)$. Therefore, it would be
interesting to apply (\ref{laggz}) to other (infrared related) problems where
such dependence on the running parameters cancels. 

\vspace{1cm} 
\noindent
{\bf Acknowledgement.} The author thanks Prof. S. Sorella, Prof. D. Zwanziger,
Dr D. Dudal, Dr C. Fischer and Dr C. McNeile for useful discussions concerning 
the Gribov problem. 

\appendix

\sect{Explicit $2$-point functions.} 

In this appendix we record the explicit forms for the one loop corrections to
each of the $2$-point functions. Using the compact notation of (\ref{twoptdef}) 
we have 
\begin{eqnarray}
X &=& -~ \left[ -~ \frac{67\pi}{192} \sqrt{C_A} \gamma^2 ~-~ 
\frac{\sqrt{4C_A \gamma^4 - (p^2)^2}}{32} \tan^{-1} \left[ -~ 
\frac{\sqrt{4C_A \gamma^4 -(p^2)^2}}{p^2} \right] \right. \nonumber \\ 
&& \left. ~~~~+~ \left[ \frac{13\sqrt{2}}{32} 
\left[ \sqrt{\left(1+\frac{16C_A\gamma^4}{(p^2)^2} \right)} - 1 \right]^{1/2} 
\ln \left[ 1 + \sqrt{1+\frac{16C_A\gamma^4}{(p^2)^2}} \right] \right. \right.
\nonumber \\
&& \left. \left. ~~~~~~~~~~-~ \frac{13\sqrt{2}}{32}  
\left[ \sqrt{\left(1+\frac{16C_A\gamma^4}{(p^2)^2} \right)} - 1 \right]^{1/2} 
\ln \left[ \frac{16C_A\gamma^4}{(p^2)^2} \right] \right. \right. \nonumber \\
&& \left. \left. ~~~~~~~~~~+~ \frac{13\sqrt{2}}{16} 
\left[ \sqrt{\left(1+\frac{16C_A\gamma^4}{(p^2)^2} \right)} - 1 \right]^{1/2} 
\ln \left[ \left( 1 + \sqrt{\left(1+\frac{16C_A\gamma^4}{(p^2)^2}\right)} 
\right)^{1/2} - \sqrt{2} \right] \right. \right. \nonumber \\
&& \left. \left. ~~~~~~~~~~+~ \frac{13\sqrt{2}}{16} 
\left[ \sqrt{\left(1+\frac{16C_A\gamma^4}{(p^2)^2} \right)} + 1 \right]^{1/2} 
\tan^{-1} \left[ \sqrt{2} \left[ \sqrt{\left(1+\frac{16C_A\gamma^4}{(p^2)^2} 
\right)} - 1 \right]^{-1/2} \right] \right. \right. \nonumber \\
&& \left. \left. ~~~~~~~~~~-~ \frac{25}{24} \tan^{-1} \left[ 
\frac{\sqrt{C_A}\gamma^2}{p^2} \right] \right] \sqrt{C_A} \gamma^2 ~-~ 
\frac{39\pi}{64} \frac{\sqrt{C_A^3}\gamma^6}{(p^2)^2} \right. \nonumber \\
&& \left. ~~~~+~ \left[ \frac{77}{96} - \frac{53}{96} \ln \left[ 1 +
\frac{(p^2)^2}{C_A\gamma^4} \right] \right] \frac{C_A\gamma^4}{p^2} ~+~ 
\frac{37}{96} \frac{\sqrt{C_A^3}\gamma^6}{(p^2)^2} \tan^{-1} \left[ 
\frac{\sqrt{C_A}\gamma^2}{p^2} \right] \right. \nonumber \\ 
&& \left. ~~~~-~ \frac{5C_A\gamma^4}{12(p^2)^2} \sqrt{4C_A\gamma^4-(p^2)^2}
\tan^{-1} \left[ - \frac{\sqrt{4C_A\gamma^4-(p^2)^2}}{p^2} \right] \right. 
\nonumber \\  
&& \left. ~~~~+~ \left[ \frac{39\sqrt{2}}{128} 
\left[ \sqrt{\left(1+\frac{16C_A\gamma^4}{(p^2)^2} \right)} + 1 \right]^{1/2} 
\ln \left[ 1 + \sqrt{1+\frac{16C_A\gamma^4}{(p^2)^2}} \right] \right. \right.
\nonumber \\
&& \left. \left. ~~~~~~~~~~-~ \frac{39\sqrt{2}}{128}  
\left[ \sqrt{\left(1+\frac{16C_A\gamma^4}{(p^2)^2} \right)} + 1 \right]^{1/2} 
\ln \left[ \frac{16C_A\gamma^4}{(p^2)^2} \right] \right. \right. \nonumber \\
&& \left. \left. ~~~~~~~~~~+~ \frac{39\sqrt{2}}{64} 
\left[ \sqrt{\left(1+\frac{16C_A\gamma^4}{(p^2)^2} \right)} + 1 \right]^{1/2} 
\ln \left[ \left( 1 + \sqrt{\left(1+\frac{16C_A\gamma^4}{(p^2)^2}\right)} 
\right)^{1/2} - \sqrt{2} \right] \right. \right. \nonumber \\
&& \left. \left. ~~~~~~~~~~-~ \frac{39\sqrt{2}}{64} 
\left[ \sqrt{\left(1+\frac{16C_A\gamma^4}{(p^2)^2} \right)} - 1 \right]^{1/2} 
\tan^{-1} \left[ \sqrt{2} \left[ \sqrt{\left(1+\frac{16C_A\gamma^4}{(p^2)^2} 
\right)} - 1 \right]^{-1/2} \right] \right. \right. \nonumber \\
&& \left. \left. ~~~~~~~~~~+~ \frac{971}{288} ~+~ \frac{163}{192} \ln \left[
\frac{(p^2)^2}{C_A\gamma^4} \right] ~+~ \frac{17}{96} \ln \left[ 
1+\frac{(p^2)^2}{C_A\gamma^4} \right] ~-~ \frac{13}{6} \ln \left[ 
\frac{p^2}{\mu^2} \right] \right] p^2 \right. \nonumber \\  
&& \left. ~~~~+~ \frac{13(p^2)^2}{384C_A\gamma^4} 
\sqrt{4C_A \gamma^4 - (p^2)^2} \tan^{-1} \left[ -~ 
\frac{\sqrt{4C_A \gamma^4 -(p^2)^2}}{p^2} \right] ~-~ 
\frac{21\pi(p^2)^2}{128\sqrt{C_A}\gamma^2} \right. \nonumber \\ 
&& \left. ~~~~+~ \left[ \frac{17\sqrt{2}}{192} 
\left[ \sqrt{\left(1+\frac{16C_A\gamma^4}{(p^2)^2} \right)} - 1 \right]^{1/2} 
\ln \left[ 1 + \sqrt{1+\frac{16C_A\gamma^4}{(p^2)^2}} \right] \right. \right.
\nonumber \\
&& \left. \left. ~~~~~~~~~~-~ \frac{17\sqrt{2}}{192}  
\left[ \sqrt{\left(1+\frac{16C_A\gamma^4}{(p^2)^2} \right)} - 1 \right]^{1/2} 
\ln \left[ \frac{16C_A\gamma^4}{(p^2)^2} \right] \right. \right. \nonumber \\
&& \left. \left. ~~~~~~~~~~+~ \frac{17\sqrt{2}}{96} 
\left[ \sqrt{\left(1+\frac{16C_A\gamma^4}{(p^2)^2} \right)} - 1 \right]^{1/2} 
\ln \left[ \left( 1 + \sqrt{\left(1+\frac{16C_A\gamma^4}{(p^2)^2}\right)} 
\right)^{1/2} - \sqrt{2} \right] \right. \right. \nonumber \\
&& \left. \left. ~~~~~~~~~~+~ \frac{17\sqrt{2}}{96} 
\left[ \sqrt{\left(1+\frac{16C_A\gamma^4}{(p^2)^2} \right)} + 1 \right]^{1/2} 
\tan^{-1} \left[ \sqrt{2} \left[ \sqrt{\left(1+\frac{16C_A\gamma^4}{(p^2)^2} 
\right)} - 1 \right]^{-1/2} \right] \right. \right. \nonumber \\
&& \left. \left. ~~~~~~~~~~+~ \frac{7}{96} \tan^{-1} \left[ 
\frac{\sqrt{C_A}\gamma^2}{p^2} \right] \right] 
\frac{(p^2)^2}{\sqrt{C_A}\gamma^2} \right. \nonumber \\ 
&& \left. ~~~~+~ \left[ \frac{5\sqrt{2}}{1536} 
\left[ \sqrt{\left(1+\frac{16C_A\gamma^4}{(p^2)^2} \right)} + 1 \right]^{1/2} 
\ln \left[ 1 + \sqrt{1+\frac{16C_A\gamma^4}{(p^2)^2}} \right] \right. \right.
\nonumber \\
&& \left. \left. ~~~~~~~~~~-~ \frac{5\sqrt{2}}{1536} 
\left[ \sqrt{\left(1+\frac{16C_A\gamma^4}{(p^2)^2} \right)} + 1 \right]^{1/2} 
\ln \left[ \frac{16C_A\gamma^4}{(p^2)^2} \right] \right. \right. \nonumber \\
&& \left. \left. ~~~~~~~~~~+~ \frac{5\sqrt{2}}{768} 
\left[ \sqrt{\left(1+\frac{16C_A\gamma^4}{(p^2)^2} \right)} + 1 \right]^{1/2} 
\ln \left[ \left( 1 + \sqrt{\left(1+\frac{16C_A\gamma^4}{(p^2)^2}\right)} 
\right)^{1/2} - \sqrt{2} \right] \right. \right. \nonumber \\
&& \left. \left. ~~~~~~~~~~-~ \frac{5\sqrt{2}}{768} 
\left[ \sqrt{\left(1+\frac{16C_A\gamma^4}{(p^2)^2} \right)} - 1 \right]^{1/2} 
\tan^{-1} \left[ \sqrt{2} \left[ \sqrt{\left(1+\frac{16C_A\gamma^4}{(p^2)^2} 
\right)} - 1 \right]^{-1/2} \right] \right. \right. \nonumber \\
&& \left. \left. ~~~~~~~~~~-~ \frac{1}{96} \ln \left[1 +
\frac{C_A\gamma^4}{(p^2)^2} \right] ~-~ \frac{1}{96} \ln \left[ 1 +
\frac{(p^2)^2}{C_A\gamma^4} \right] \right] \frac{(p^2)^3}{C_A\gamma^4} 
\right] C_A a \nonumber \\  
&& -~ \left[ \frac{4}{3} \ln \left( \frac{p^2}{\mu^2} \right) 
- \frac{20}{9} \right] p^2 T_F \Nf a ~+~ O(a^2) ~. 
\end{eqnarray} 
For the mixed propagator we have $M$ $=$ $U$ with 
\begin{eqnarray}
U &=& -~ \left[ \left[ \frac{1}{64} \ln \left( 1 + \frac{(p^2)^2}{C_A\gamma^4} 
\right) - \frac{31}{64} \right] ~-~ \frac{C_A\gamma^4}{96(p^2)^2} \ln 
\left[ 1 + \frac{(p^2)^2}{C_A\gamma^4} \right] ~+~ 
\frac{179\pi\sqrt{C_A}\gamma^2}{384p^2} \right. \nonumber \\
&& \left. ~~~~-~ \frac{11\sqrt{C_A}\gamma^2}{192 p^2} \tan^{-1} \left[ 
\frac{\sqrt{C_A}\gamma^2}{p^2} \right] ~+~ 
\frac{7\sqrt{4C_A \gamma^4 - (p^2)^2}}{16p^2} \tan^{-1} \left[ -~ 
\frac{\sqrt{4C_A \gamma^4 -(p^2)^2}}{p^2} \right] \right. \nonumber \\ 
&& \left. ~~~~-~ \frac{7 p^2 \sqrt{4C_A \gamma^4 - (p^2)^2}}{64C_A\gamma^4} 
\tan^{-1} \left[ -~ \frac{\sqrt{4C_A \gamma^4 -(p^2)^2}}{p^2} \right]  
- \frac{39\pi p^2}{128\sqrt{C_A}\gamma^2}
- \frac{\pi(p^2)^3}{256\sqrt{C_A^3}\gamma^6} \right. \nonumber \\
&& \left. ~~~~+~ \left[ \frac{7\sqrt{2}}{64} 
\left[ \sqrt{\left(1+\frac{16C_A\gamma^4}{(p^2)^2} \right)} - 1 \right]^{1/2} 
\ln \left[ 1 + \sqrt{1+\frac{16C_A\gamma^4}{(p^2)^2}} \right] \right.
\right. \nonumber \\
&& \left. \left. ~~~~~~~~~~-~ \frac{7\sqrt{2}}{64}  
\left[ \sqrt{\left(1+\frac{16C_A\gamma^4}{(p^2)^2} \right)} - 1 \right]^{1/2} 
\ln \left[ \frac{16C_A\gamma^4}{(p^2)^2} \right] \right. \right. \nonumber \\
&& \left. \left. ~~~~~~~~~~+~ \frac{7\sqrt{2}}{32} 
\left[ \sqrt{\left(1+\frac{16C_A\gamma^4}{(p^2)^2} \right)} - 1 \right]^{1/2} 
\ln \left[ \left( 1 + \sqrt{\left(1+\frac{16C_A\gamma^4}{(p^2)^2}\right)} 
\right)^{1/2} - \sqrt{2} \right] \right. \right. \nonumber \\
&& \left. \left. ~~~~~~~~~~+~ \frac{7\sqrt{2}}{32} 
\left[ \sqrt{\left(1+\frac{16C_A\gamma^4}{(p^2)^2} \right)} + 1 \right]^{1/2} 
\tan^{-1} \left[ \sqrt{2} \left[ \sqrt{\left(1+\frac{16C_A\gamma^4}{(p^2)^2} 
\right)} - 1 \right]^{-1/2} \right] \right. \right. \nonumber \\
&& \left. \left. ~~~~~~~~~~-~ \frac{1}{24} \tan^{-1} \left[ 
\frac{\sqrt{C_A}\gamma^2}{p^2} \right] \right] \frac{p^2}{\sqrt{C_A}\gamma^2} 
\right. \nonumber \\ 
&& \left. ~~~~+~ \left[ -~ \frac{5\sqrt{2}}{256} 
\left[ \sqrt{\left(1+\frac{16C_A\gamma^4}{(p^2)^2} \right)} + 1 \right]^{1/2} 
\ln \left[ 1 + \sqrt{1+\frac{16C_A\gamma^4}{(p^2)^2}} \right] \right.
\right. \nonumber \\
&& \left. \left. ~~~~~~~~~~+~ \frac{5\sqrt{2}}{256}  
\left[ \sqrt{\left(1+\frac{16C_A\gamma^4}{(p^2)^2} \right)} + 1 \right]^{1/2} 
\ln \left[ \frac{16C_A\gamma^4}{(p^2)^2} \right] \right. \right. \nonumber \\
&& \left. \left. ~~~~~~~~~~-~ \frac{5\sqrt{2}}{128} 
\left[ \sqrt{\left(1+\frac{16C_A\gamma^4}{(p^2)^2} \right)} - 1 \right]^{1/2} 
\ln \left[ \left( 1 + \sqrt{\left(1+\frac{16C_A\gamma^4}{(p^2)^2}\right)} 
\right)^{1/2} - \sqrt{2} \right] \right. \right. \nonumber \\
&& \left. \left. ~~~~~~~~~~+~ \frac{5\sqrt{2}}{128} 
\left[ \sqrt{\left(1+\frac{16C_A\gamma^4}{(p^2)^2} \right)} + 1 \right]^{1/2} 
\tan^{-1} \left[ \sqrt{2} \left[ \sqrt{\left(1+\frac{16C_A\gamma^4}{(p^2)^2} 
\right)} - 1 \right]^{-1/2} \right] \right. \right. \nonumber \\
&& \left. \left. ~~~~~~~~~~+~ \frac{5}{192} \ln 
\left( 1 + \frac{(p^2)^2}{C_A\gamma^4} \right) - \frac{1}{384} \ln \left( 
\frac{(p^2)^2}{C_A\gamma^4} \right) \right] \frac{(p^2)^2}{C_A\gamma^4} \right.
\nonumber \\ 
&& \left. ~~~~+~ \left[ \frac{\sqrt{2}}{512} 
\left[ \sqrt{\left(1+\frac{16C_A\gamma^4}{(p^2)^2} \right)} - 1 \right]^{1/2} 
\ln \left[ 1 + \sqrt{1+\frac{16C_A\gamma^4}{(p^2)^2}} \right] \right.
\right. \nonumber \\
&& \left. \left. ~~~~~~~~~~-~ \frac{\sqrt{2}}{512}  
\left[ \sqrt{\left(1+\frac{16C_A\gamma^4}{(p^2)^2} \right)} - 1 \right]^{1/2} 
\ln \left[ \frac{16C_A\gamma^4}{(p^2)^2} \right] \right. \right. 
\nonumber \\
&& \left. \left. ~~~~~~~~~~+~ \frac{\sqrt{2}}{256} 
\left[ \sqrt{\left(1+\frac{16C_A\gamma^4}{(p^2)^2} \right)} - 1 \right]^{1/2} 
\ln \left[ \left( 1 + \sqrt{\left(1+\frac{16C_A\gamma^4}{(p^2)^2}\right)} 
\right)^{1/2} - \sqrt{2} \right] \right. \right. \nonumber \\
&& \left. \left. ~~~~~~~~~~+~ \frac{\sqrt{2}}{256} 
\left[ \sqrt{\left(1+\frac{16C_A\gamma^4}{(p^2)^2} \right)} + 1 \right]^{1/2} 
\tan^{-1} \left[ \sqrt{2} \left[ \sqrt{\left(1+\frac{16C_A\gamma^4}{(p^2)^2} 
\right)} - 1 \right]^{-1/2} \right] \right. \right. \nonumber \\
&& \left. \left. ~~~~~~~~~~+~ \frac{1}{64} \tan^{-1} 
\left[ \frac{\sqrt{C_A}\gamma^2}{p^2} 
\right] \right] \frac{(p^2)^3}{\sqrt{C_A^3}\gamma^6} \right] C_A \gamma^2 a ~+~
O(a^2)  
\end{eqnarray} 
The various colour channels of the $\phi^{ab}_\mu$ $2$-point function are
\begin{eqnarray}
Q &=& -~ \left[ -~ \frac{3\pi}{8} \sqrt{C_A} \gamma^2 ~+~ \frac{3}{4}
\sqrt{C_A} \gamma^2 \tan^{-1} \left[ \frac{\sqrt{C_A}\gamma^2}{p^2} \right] ~-~ 
\frac{(p^2)^2}{4\sqrt{C_A}\gamma^2} \tan^{-1} \left[ 
\frac{\sqrt{C_A}\gamma^2}{p^2} \right] \right. \nonumber \\
&& \left. ~~~~+~ \frac{C_A \gamma^4}{8p^2} \ln 
\left[ 1 + \frac{(p^2)^2}{C_A\gamma^4} \right] ~+~ \left[ \frac{5}{4} ~-~ 
\frac{3}{8} \ln \left( \frac{[(p^2)^2 + C_A \gamma^4]}{\mu^4} \right) 
\right] p^2 \right] C_A a \nonumber \\
&& +~ O(a^2) 
\end{eqnarray}  
which is proportional to $\left(D_c(p^2)\right)^{-1}$ 
\begin{eqnarray}
W &=& -~ \left[ \frac{25\pi}{576} \sqrt{C_A} \gamma^2 ~-~ 
\frac{17}{144} \sqrt{C_A} \gamma^2 \tan^{-1} \left[ 
\frac{\sqrt{C_A}\gamma^2}{p^2} \right] \right. \nonumber \\
&& \left. ~~~~-~ \frac{\sqrt{4C_A \gamma^4 - (p^2)^2}}{72} \tan^{-1} \left[ -~ 
\frac{\sqrt{4C_A \gamma^4 -(p^2)^2}}{p^2} \right] \right. \nonumber \\ 
&& \left. ~~~~+~ \frac{C_A\gamma^4}{72(p^2)^2} \sqrt{4C_A \gamma^4 - (p^2)^2} 
\tan^{-1} \left[ -~ \frac{\sqrt{4C_A \gamma^4 -(p^2)^2}}{p^2} \right] \right. 
\nonumber \\ 
&& \left. ~~~~+~ \frac{\pi\sqrt{C_A^3}\gamma^6}{144(p^2)^2} ~+~ 
\frac{1}{72} \frac{\sqrt{C_A^3}\gamma^6}{(p^2)^2} \tan^{-1} \left[ 
\frac{\sqrt{C_A}\gamma^2}{p^2} \right] ~-~ \frac{19C_A\gamma^4}{576p^2} \ln
\left[ 1 + \frac{(p^2)^2}{C_A\gamma^4} \right] \right. \nonumber \\
&& \left. ~~~~+~ \left[ \frac{\sqrt{2}}{72} 
\left[ \sqrt{\left(1+\frac{16C_A\gamma^4}{(p^2)^2} \right)} + 1 \right]^{1/2} 
\ln \left[ 1 + \sqrt{1+\frac{16C_A\gamma^4}{(p^2)^2}} \right] \right. \right.
\nonumber \\
&& \left. \left. ~~~~~~~~~~-~ \frac{\sqrt{2}}{72}  
\left[ \sqrt{\left(1+\frac{16C_A\gamma^4}{(p^2)^2} \right)} + 1 \right]^{1/2} 
\ln \left[ \frac{16C_A\gamma^4}{(p^2)^2} \right] \right. \right. \nonumber \\
&& \left. \left. ~~~~~~~~~~+~ \frac{\sqrt{2}}{36} 
\left[ \sqrt{\left(1+\frac{16C_A\gamma^4}{(p^2)^2} \right)} + 1 \right]^{1/2} 
\ln \left[ \left( 1 + \sqrt{\left(1+\frac{16C_A\gamma^4}{(p^2)^2}\right)} 
\right)^{1/2} - \sqrt{2} \right] \right. \right. \nonumber \\
&& \left. \left. ~~~~~~~~~~-~ \frac{\sqrt{2}}{36} 
\left[ \sqrt{\left(1+\frac{16C_A\gamma^4}{(p^2)^2} \right)} - 1 \right]^{1/2} 
\tan^{-1} \left[ \sqrt{2} \left[ \sqrt{\left(1+\frac{16C_A\gamma^4}{(p^2)^2} 
\right)} - 1 \right]^{-1/2} \right] \right. \right. \nonumber \\
&& \left. \left. ~~~~~~~~~~-~ \frac{1}{96} ~+~ \frac{7}{144} \ln \left[ 1 
+ \frac{(p^2)^2}{C_A\gamma^4} \right] \right] p^2 \right. \nonumber \\  
&& \left. ~~~~+~ \frac{(p^2)^2}{384C_A\gamma^4} \sqrt{4C_A \gamma^4 - (p^2)^2} 
\tan^{-1} \left[ -~ \frac{\sqrt{4C_A \gamma^4 -(p^2)^2}}{p^2} \right] ~-~ 
\frac{5\pi(p^2)^2}{1152\sqrt{C_A}\gamma^2} \right. \nonumber \\ 
&& \left. ~~~~+~ \left[ \frac{\sqrt{2}}{384} 
\left[ \sqrt{\left(1+\frac{16C_A\gamma^4}{(p^2)^2} \right)} - 1 \right]^{1/2} 
\ln \left[ 1 + \sqrt{1+\frac{16C_A\gamma^4}{(p^2)^2}} \right] \right. \right.
\nonumber \\
&& \left. \left. ~~~~~~~~~~-~ \frac{\sqrt{2}}{384} 
\left[ \sqrt{\left(1+\frac{16C_A\gamma^4}{(p^2)^2} \right)} - 1 \right]^{1/2} 
\ln \left[ \frac{16C_A\gamma^4}{(p^2)^2} \right] \right. \right. \nonumber \\
&& \left. \left. ~~~~~~~~~~+~ \frac{\sqrt{2}}{192} 
\left[ \sqrt{\left(1+\frac{16C_A\gamma^4}{(p^2)^2} \right)} - 1 \right]^{1/2} 
\ln \left[ \left( 1 + \sqrt{\left(1+\frac{16C_A\gamma^4}{(p^2)^2}\right)} 
\right)^{1/2} - \sqrt{2} \right] \right. \right. \nonumber \\
&& \left. \left. ~~~~~~~~~~+~ \frac{\sqrt{2}}{192} 
\left[ \sqrt{\left(1+\frac{16C_A\gamma^4}{(p^2)^2} \right)} + 1 \right]^{1/2} 
\tan^{-1} \left[ \sqrt{2} \left[ \sqrt{\left(1+\frac{16C_A\gamma^4}{(p^2)^2} 
\right)} - 1 \right]^{-1/2} \right] \right. \right. \nonumber \\
&& \left. \left. ~~~~~~~~~~+~ \frac{5}{144} \tan^{-1} \left[ 
\frac{\sqrt{C_A}\gamma^2}{p^2} \right] \right] 
\frac{(p^2)^2}{\sqrt{C_A}\gamma^2} \right. \nonumber \\  
&& \left. ~~~~+~ \left[ \frac{\sqrt{2}}{4608} 
\left[ \sqrt{\left(1+\frac{16C_A\gamma^4}{(p^2)^2} \right)} + 1 \right]^{1/2} 
\ln \left[ 1 + \sqrt{1+\frac{16C_A\gamma^4}{(p^2)^2}} \right] \right. \right.
\nonumber \\
&& \left. \left. ~~~~~~~~~~-~ \frac{\sqrt{2}}{4608} 
\left[ \sqrt{\left(1+\frac{16C_A\gamma^4}{(p^2)^2} \right)} + 1 \right]^{1/2} 
\ln \left[ \frac{16C_A\gamma^4}{(p^2)^2} \right] \right. \right. \nonumber \\
&& \left. \left. ~~~~~~~~~~+~ \frac{\sqrt{2}}{2304} 
\left[ \sqrt{\left(1+\frac{16C_A\gamma^4}{(p^2)^2} \right)} + 1 \right]^{1/2} 
\ln \left[ \left( 1 + \sqrt{\left(1+\frac{16C_A\gamma^4}{(p^2)^2}\right)} 
\right)^{1/2} - \sqrt{2} \right] \right. \right. \nonumber \\
&& \left. \left. ~~~~~~~~~~-~ \frac{\sqrt{2}}{2304} 
\left[ \sqrt{\left(1+\frac{16C_A\gamma^4}{(p^2)^2} \right)} - 1 \right]^{1/2} 
\tan^{-1} \left[ \sqrt{2} \left[ \sqrt{\left(1+\frac{16C_A\gamma^4}{(p^2)^2} 
\right)} - 1 \right]^{-1/2} \right] \right. \right. \nonumber \\
&& \left. \left. ~~~~~~~~~~-~ \frac{1}{1152} \ln \left[ 
\frac{(p^2)^2}{C_A\gamma^4} \right] ~-~ \frac{1}{576} \ln \left[ 1 +
\frac{C_A\gamma^4}{(p^2)^2} \right] \right] \frac{(p^2)^3}{C_A\gamma^4}
\right] a ~+~ O(a^2)  
\end{eqnarray}  
\begin{eqnarray}
R &=& -~ \left[ \frac{7\pi}{144} \sqrt{C_A} \gamma^2 ~-~ 
\frac{5}{36} \sqrt{C_A} \gamma^2 \tan^{-1} \left[ 
\frac{\sqrt{C_A}\gamma^2}{p^2} \right] \right. \nonumber \\
&& \left. ~~~~-~ \frac{\sqrt{4C_A \gamma^4 - (p^2)^2}}{72} \tan^{-1} \left[ -~ 
\frac{\sqrt{4C_A \gamma^4 -(p^2)^2}}{p^2} \right] \right. \nonumber \\ 
&& \left. ~~~~+~ \frac{C_A\gamma^4}{18(p^2)^2} \sqrt{4C_A \gamma^4 - (p^2)^2} 
\tan^{-1} \left[ -~ \frac{\sqrt{4C_A \gamma^4 -(p^2)^2}}{p^2} \right] ~-~ 
\frac{C_A\gamma^4}{32p^2} \right. \nonumber \\ 
&& \left. ~~~~+~ \frac{25\pi\sqrt{C_A^3}\gamma^6}{576(p^2)^2} ~+~ 
\frac{7}{288} \frac{\sqrt{C_A^3}\gamma^6}{(p^2)^2} \tan^{-1} \left[ 
\frac{\sqrt{C_A}\gamma^2}{p^2} \right] ~-~ \frac{7C_A\gamma^4}{144p^2} \ln 
\left[ 1 + \frac{(p^2)^2}{C_A\gamma^4} \right] \right. \nonumber \\
&& \left. ~~~~+~ \left[ \frac{\sqrt{2}}{72} 
\left[ \sqrt{\left(1+\frac{16C_A\gamma^4}{(p^2)^2} \right)} + 1 \right]^{1/2} 
\ln \left[ 1 + \sqrt{1+\frac{16C_A\gamma^4}{(p^2)^2}} \right] \right. \right.
\nonumber \\
&& \left. \left. ~~~~~~~~~~-~ \frac{\sqrt{2}}{72}  
\left[ \sqrt{\left(1+\frac{16C_A\gamma^4}{(p^2)^2} \right)} + 1 \right]^{1/2} 
\ln \left[ \frac{16C_A\gamma^4}{(p^2)^2} \right] \right. \right. \nonumber \\
&& \left. \left. ~~~~~~~~~~+~ \frac{\sqrt{2}}{36} 
\left[ \sqrt{\left(1+\frac{16C_A\gamma^4}{(p^2)^2} \right)} + 1 \right]^{1/2} 
\ln \left[ \left( 1 + \sqrt{\left(1+\frac{16C_A\gamma^4}{(p^2)^2}\right)} 
\right)^{1/2} - \sqrt{2} \right] \right. \right. \nonumber \\
&& \left. \left. ~~~~~~~~~~-~ \frac{\sqrt{2}}{36} 
\left[ \sqrt{\left(1+\frac{16C_A\gamma^4}{(p^2)^2} \right)} - 1 \right]^{1/2} 
\tan^{-1} \left[ \sqrt{2} \left[ \sqrt{\left(1+\frac{16C_A\gamma^4}{(p^2)^2} 
\right)} - 1 \right]^{-1/2} \right] \right. \right. \nonumber \\
&& \left. \left. ~~~~~~~~~~-~ \frac{1}{96} ~+~ \frac{11}{288} \ln \left[ 1 
+ \frac{(p^2)^2}{C_A\gamma^4} \right] \right] p^2 \right. \nonumber \\  
&& \left. ~~~~+~ \frac{\pi(p^2)^2}{288\sqrt{C_A}\gamma^2} ~+~ 
\frac{(p^2)^2}{288\sqrt{C_A}\gamma^2} \tan^{-1} \left[ 
\frac{\sqrt{C_A}\gamma^2}{p^2} \right] \right. \nonumber \\  
&& \left. ~~~~+~ \left[ \frac{\sqrt{2}}{1152} 
\left[ \sqrt{\left(1+\frac{16C_A\gamma^4}{(p^2)^2} \right)} + 1 \right]^{1/2} 
\ln \left[ 1 + \sqrt{1+\frac{16C_A\gamma^4}{(p^2)^2}} \right] \right. \right.
\nonumber \\
&& \left. \left. ~~~~~~~~~~-~ \frac{\sqrt{2}}{1152} 
\left[ \sqrt{\left(1+\frac{16C_A\gamma^4}{(p^2)^2} \right)} + 1 \right]^{1/2} 
\ln \left[ \frac{16C_A\gamma^4}{(p^2)^2} \right] \right. \right. \nonumber \\
&& \left. \left. ~~~~~~~~~~+~ \frac{\sqrt{2}}{576} 
\left[ \sqrt{\left(1+\frac{16C_A\gamma^4}{(p^2)^2} \right)} + 1 \right]^{1/2} 
\ln \left[ \left( 1 + \sqrt{\left(1+\frac{16C_A\gamma^4}{(p^2)^2}\right)} 
\right)^{1/2} - \sqrt{2} \right] \right. \right. \nonumber \\
&& \left. \left. ~~~~~~~~~~-~ \frac{\sqrt{2}}{576} 
\left[ \sqrt{\left(1+\frac{16C_A\gamma^4}{(p^2)^2} \right)} - 1 \right]^{1/2} 
\tan^{-1} \left[ \sqrt{2} \left[ \sqrt{\left(1+\frac{16C_A\gamma^4}{(p^2)^2} 
\right)} - 1 \right]^{-1/2} \right] \right. \right. \nonumber \\
&& \left. \left. ~~~~~~~~~~+~ \frac{1}{576} \ln \left[ 
\frac{(p^2)^2}{C_A\gamma^4} \right] ~-~ \frac{1}{288} \ln \left[ 1 +
\frac{C_A\gamma^4}{(p^2)^2} \right] \right] \frac{(p^2)^3}{C_A\gamma^4}
\right] a ~+~ O(a^2)  
\end{eqnarray}
and
\begin{eqnarray}
S &=& -~ \left[ -~ \frac{\sqrt{4C_A \gamma^4 - (p^2)^2}}{12p^2} \tan^{-1} 
\left[ -~ \frac{\sqrt{4C_A \gamma^4 -(p^2)^2}}{p^2} \right] ~-~  
\frac{17}{24} \sqrt{C_A}\gamma^2 \tan^{-1} \left[ 
\frac{\sqrt{C_A}\gamma^2}{p^2} \right] \right. \nonumber \\
&& \left. ~~~+~ \frac{25\pi}{96} \sqrt{C_A} \gamma^2 ~+~ 
\frac{C_A\gamma^4}{12(p^2)^2} \sqrt{4C_A \gamma^4 - (p^2)^2} \tan^{-1} 
\left[ -~ \frac{\sqrt{4C_A \gamma^4 -(p^2)^2}}{p^2} \right] \right.
\nonumber \\
&& \left. ~~~+~ \frac{1}{12} \frac{\sqrt{C_A^3}\gamma^6}{(p^2)^2} \tan^{-1} 
\left[ \frac{\sqrt{C_A}\gamma^2}{p^2} \right] ~+~
\frac{\pi}{24} \frac{\sqrt{C_A^3}\gamma^6}{(p^2)^2} ~-~ 
\frac{19C_A\gamma^4}{96p^2} \ln \left[ 1 + \frac{(p^2)^2}{C_A\gamma^4}
\right] \right. \nonumber \\
&& \left. ~~+~ \left[ -~ \frac{1}{16} + \frac{7}{24} \ln \left[ 1 +
\frac{C_A\gamma^4}{(p^2)^2} \right] \right. \right. \nonumber \\ 
&& \left. \left. ~~+~ \frac{\sqrt{2}}{12} 
\left[ \sqrt{\left(1+\frac{16C_A\gamma^4}{(p^2)^2} \right)} + 1 \right]^{1/2} 
\ln \left[ 1 + \sqrt{1+\frac{16C_A\gamma^4}{(p^2)^2}} \right] \right. \right. 
\nonumber \\
&& \left. \left. ~~~~~~~~~~-~ \frac{\sqrt{2}}{12}  
\left[ \sqrt{\left(1+\frac{16C_A\gamma^4}{(p^2)^2} \right)} + 1 \right]^{1/2} 
\ln \left[ \frac{16C_A\gamma^4}{(p^2)^2} \right] \right. \right. \nonumber \\
&& \left. \left. ~~~~~~~~~~+~ \frac{\sqrt{2}}{6} 
\left[ \sqrt{\left(1+\frac{16C_A\gamma^4}{(p^2)^2} \right)} + 1 \right]^{1/2} 
\ln \left[ \left( 1 + \sqrt{\left(1+\frac{16C_A\gamma^4}{(p^2)^2}\right)} 
\right)^{1/2} - \sqrt{2} \right] \right. \right. \nonumber \\
&& \left. \left. ~~~~~~~~~~-~ \frac{\sqrt{2}}{6} 
\left[ \sqrt{\left(1+\frac{16C_A\gamma^4}{(p^2)^2} \right)} - 1 \right]^{1/2} 
\tan^{-1} \left[ \sqrt{2} \left[ \sqrt{\left(1+\frac{16C_A\gamma^4}{(p^2)^2} 
\right)} - 1 \right]^{-1/2} \right] p^2 \right. \right. \nonumber \\
&& \left. ~~+~ \left[ \frac{\sqrt{2}}{64} 
\left[ \sqrt{\left(1+\frac{16C_A\gamma^4}{(p^2)^2} \right)} - 1 \right]^{1/2} 
\ln \left[ 1 + \sqrt{1+\frac{16C_A\gamma^4}{(p^2)^2}} \right] \right. \right.
\nonumber \\
&& \left. \left. ~~~~~~~~~~-~ \frac{\sqrt{2}}{64}  
\left[ \sqrt{\left(1+\frac{16C_A\gamma^4}{(p^2)^2} \right)} - 1 \right]^{1/2} 
\ln \left[ \frac{16C_A\gamma^4}{(p^2)^2} \right] \right. \right. \nonumber \\
&& \left. \left. ~~~~~~~~~~+~ \frac{\sqrt{2}}{32} 
\left[ \sqrt{\left(1+\frac{16C_A\gamma^4}{(p^2)^2} \right)} - 1 \right]^{1/2} 
\ln \left[ \left( 1 + \sqrt{\left(1+\frac{16C_A\gamma^4}{(p^2)^2}\right)} 
\right)^{1/2} - \sqrt{2} \right] \right. \right. \nonumber \\
&& \left. \left. ~~~~~~~~~~+~ \frac{\sqrt{2}}{32} 
\left[ \sqrt{\left(1+\frac{16C_A\gamma^4}{(p^2)^2} \right)} + 1 \right]^{1/2} 
\tan^{-1} \left[ \sqrt{2} \left[ \sqrt{\left(1+\frac{16C_A\gamma^4}{(p^2)^2} 
\right)} - 1 \right]^{-1/2} \right] \right. \right. \nonumber \\
&& \left. \left. ~~~~~~~~~~+~ \frac{5}{24} \tan^{-1} \left[ 
\frac{\sqrt{C_A}\gamma^2}{p^2} \right] \right] 
\frac{(p^2)^2}{\sqrt{C_A}\gamma^2} \right. \nonumber \\
&& \left. ~~+~ \frac{(p^2)^2}{64C_A\gamma^4} 
\frac{\sqrt{4C_A \gamma^4 - (p^2)^2}}{12p^2} \tan^{-1} 
\left[ -~ \frac{\sqrt{4C_A \gamma^4 -(p^2)^2}}{p^2} \right] ~-~ 
\frac{5(p^2)^2\pi}{192\sqrt{C_A}\gamma^2} \right. \nonumber \\ 
&& \left. ~~+~ \left[ \frac{\sqrt{2}}{768} 
\left[ \sqrt{\left(1+\frac{16C_A\gamma^4}{(p^2)^2} \right)} + 1 \right]^{1/2} 
\ln \left[ 1 + \sqrt{1+\frac{16C_A\gamma^4}{(p^2)^2}} \right] \right. \right.
\nonumber \\
&& \left. \left. ~~~~~~~-~ \frac{\sqrt{2}}{768}  
\left[ \sqrt{\left(1+\frac{16C_A\gamma^4}{(p^2)^2} \right)} + 1 \right]^{1/2} 
\ln \left[ \frac{16C_A\gamma^4}{(p^2)^2} \right] \right. \right. \nonumber \\
&& \left. \left. ~~~~~~~+~ \frac{\sqrt{2}}{384} 
\left[ \sqrt{\left(1+\frac{16C_A\gamma^4}{(p^2)^2} \right)} + 1 \right]^{1/2} 
\ln \left[ \left( 1 + \sqrt{\left(1+\frac{16C_A\gamma^4}{(p^2)^2}\right)} 
\right)^{1/2} - \sqrt{2} \right] \right. \right. \nonumber \\
&& \left. \left. ~~~~~~~-~ \frac{\sqrt{2}}{384} 
\left[ \sqrt{\left(1+\frac{16C_A\gamma^4}{(p^2)^2} \right)} - 1 \right]^{1/2} 
\tan^{-1} \left[ \sqrt{2} \left[ \sqrt{\left(1+\frac{16C_A\gamma^4}{(p^2)^2} 
\right)} - 1 \right]^{-1/2} \right] \right. \right. \nonumber \\
&& \left. \left. ~~+~ \frac{1}{192} \ln \left[ 
\frac{C_A\gamma^4}{(p^2)^2} \right] ~-~ \frac{1}{96} \ln \left[ 
1 + \frac{C_A\gamma^4}{(p^2)^2} \right] \right] \frac{(p^2)^3}{C_A\gamma^4}  
\right] \frac{a}{C_A} ~+~ O(a^2) ~. 
\end{eqnarray} 
We note that in these expressions the explicit $\mu$ dependence appears only
in a small set of terms, which can clearly be identified with the piece of the
$2$-point function deriving from the ultraviolet contribution.

\end{document}